\documentclass[useAMS,usenatbib]{mn2e}

\usepackage{xspace}
\usepackage{graphicx}
\usepackage{color}

\newcommand{\isoage}{13}
\newcommand{\kms}{\mathrm{km~s^{-1}}}
\newcommand{\Teff}{\mathrm{T_{eff}}}
\newcommand{\Msun}{\mathrm{M_{\odot}}}
\newcommand{\Lsun}{\mathrm{L_{\odot}}}
\newcommand{\dY}{\Delta\mathrm{Y}}
\newcommand{\CdY}{\mathrm{C}\Delta\mathrm{Y}}
\newcommand{\phx}{{\sc Phoenix}}
\newcommand{\comment}[1]{ #1 }

\title[Stellar Models of Multiple Populations]{Stellar Models of Multiple Populations in Globular Clusters. 
I. The Main Sequence of NGC\,6752}
\author[Dotter et al.]{Aaron Dotter$^{1}$\thanks{email: aaron.dotter@gmail.com}, 
Jason W.\ Ferguson$^{2}$, Charlie Conroy$^3$,
\newauthor
 A.~P.\ Milone$^1$, A.~F.\ Marino$^1$, and David Yong$^1$\\
$^{1}$Research School of Astronomy and Astrophysics, Australian National University, Canberra, ACT, Australia\\
$^{2}$Department of Mathematics, Statistics, and Physics, Wichita State University, Wichita, KS, USA \\
$^3$Department of Astrnomy and Astrophysics, University of California, Santa Cruz, CA, USA}
\begin{document}

\date{Accepted 2014 October 16. Received 2014 October 15; in original form 2014 August 14}

\pagerange{\pageref{firstpage}--\pageref{lastpage}} \pubyear{2014}

\maketitle

\label{firstpage}

\begin{abstract}
We present stellar atmosphere and evolution models of main sequence stars in two stellar 
populations of the Galactic globular cluster NGC\,6752. These populations represent the two extremes of
light-element abundance variations in the cluster. NGC\,6752 is a benchmark cluster in the study of multiple 
stellar populations because of the rich array of spectroscopic abundances and panchromatic \emph{Hubble Space 
Telescope} photometry.
The spectroscopic abundances are used to compute stellar atmosphere and evolution models. 
The synthetic spectra for the two populations show significant differences in the ultraviolet 
and, for the coolest temperatures, in the near-infrared.  The stellar evolution models exhibit insignificant 
differences in the H-R diagram except on the lower main sequence. The appearance of multiple sequences in the 
colour-magnitude diagrams (CMDs) of NGC\,6752 is almost exclusively due to spectral effects caused by the 
abundance variations. The models reproduce the observed splitting and/or broadening of sequences in a range of CMDs.
The ultraviolet CMDs are sensitive to variations in carbon, nitrogen, and oxygen but the models are not reliable 
enough to directly estimate abundance variations from photometry. On the other hand, the widening of the lower 
main sequence in the near-infrared CMD, driven by oxygen-variation via the water molecule, is well-described by 
the models and can be used to estimate the range of oxygen present in a cluster from photometry.
We confirm that it is possible to use multiband photometry to estimate helium variations among the different 
populations, with the caveat that the estimated amount of helium-enhancement is model-dependent.
\end{abstract}


\section{Introduction}
The study of chemical abundance variations in globular clusters (GCs) has a long history, the essence 
of which is captured in reviews by \citet{FreemanNorris1981}, \citet{Gratton2004}, and \citet{Gratton2012}. 
Recent work has vastly increased the number of GCs in which such variations are observed 
as well as the sample size in a given cluster \citep[e.g.,][and so on]{Yong2008,Marino2008,Carretta2009A,Carretta2009B,Marino2011}. 
These studies highlight the 
spread in light elements, especially the well-established anticorrelations between oxygen and sodium and, to a lesser 
extent, magnesium and aluminum. These variations are connected with the products of nuclear burning at temperatures appropriate
for the hot CNO cycle \citep[e.g.,][]{Prantzos2007}, in intermediate- and high-mass stars, though the exact source of 
the nucleosynthesis site and nature of the pollution mechanism remain the subject of debate.

\comment{Identifying distinct sequences within a single GC solely from photometry began with the red giant branch 
\citep[RGB;][]{Lee1999} and main sequence \citep[MS;][]{JayPhD,Bedin2004} of $\omega$ Centauri.} 
Photometric discovery of multiple sequences has accelerated with the sensitivity and 
resolution of detectors on board the \emph{Hubble Space Telescope (HST)}, see \citet{Piotto2009} for 
a review of discoveries up until that time. The UVIS and IR channels of the \emph{HST} Wide Field Camera 
3 (WFC3) are the most effective means of detecting multiple photometric sequences 
available at present \citep[e.g.,][]{Milone2013}. 

The link between multiple stellar sequences in the colour-magnitude diagram (CMD), as observed photometrically, and 
light-element abundance variations, as observed spectroscopically, is an area of active research. Meanwhile, 
there have been a handful of studies that model the appearance of multiple stellar populations in GCs. 
The first major work on the influence of light-element adundance variations on stellar evolution models 
is that of \citet{Salaris2006}, who compared stellar evolution models with a typical $\alpha$-enhanced abundance 
pattern (oxygen-rich) to models with a composition reflective of the extreme end of the observed abundance 
variations (oxygen-poor). Salaris et al.\ found that the abundance variations produce quite a small spread 
in effective temperature ($\Teff$) for coeval populations and, therefore, CMDs
comprising broadband, optical filters could only show large spreads in the presence of a significant variation in helium.
\citet{Pietrinferni2009} expanded the BaSTI database of stellar evolution models to include light-element variations 
following the same abundance pattern as \citet{Salaris2006}, with masses from 0.4 to $1.2 \Msun$.

\citet{Sbordone2011} made the first systematic comparison of model atmospheres and synthetic spectra from 3,000 to 
10,000\AA\ computed with the abundance patterns of \citet{Salaris2006}. Models were computed for [Fe/H]=$-1.62$ at 
8 different $\Teff$-log(g) pairs corresponding to MS, subgiant, and red giant stars in the BaSTI isochrones. 
The authors found modest differences in the atmosphere structures due to variations in the light elements. On the other hand, 
the synthetic spectra and associated colour transformations were significantly influenced by differences in the absorption 
of carbon-, nitrogen-, and oxygen-bearing molecules at wavelengths shorter than 4,000\AA, while leaving longer wavelengths 
essentially unchanged; the near-infrared was not considered. 

Sbordone et al.\ recommend broadband U and B filters, as well as Str\"omgren u, v, and y,
to maximise the separation of GC stars in the CMD. The authors found that increasing the 
helium content made essentially no change to the model atmospheres and synthetic spectra and, thus, its influence is restricted 
to the interior models \citep[in agreement with][]{Girardi2007}. \citet{Sbordone2011} found the separation of the sequences 
in a variety of CMDs to be consistent with observations but made no direct comparison between their models and photometry.

\citet{diCriscienzo2010A} used the \emph{HST} photometry of NGC\,6397 \citep{Richer2008} and predictions for the enhancement 
of helium, carbon, nitrogen, and oxygen at the metallicity of the cluster to estimate the allowed spread in helium of 2\% by mass ($\dY=0.02$) 
assuming C+N+O remains roughly constant or as much as 4\% ($\dY=0.04$) if C+N+O is allowed to increase along with helium. The 
photometry used in this case was limited to the $F606W$ and $F814W$ bands that are insensitive to abundance variations
\citep[see][and Sections \ref{atm} and \ref{dsep} of this paper]{Sbordone2011}.

This paper presents a case study of the Galactic GC NGC\,6752 that, due to its proximity and
other distinguishing characteristics, has been the focus of significant observational effort in both spectroscopy
\citep{Grundahl2002,Yong2003,Yong2005,Yong2008,Yong2013} and photometry \citep[][]{Milone2010,Milone2013}. 
These complimentary data sets invite a careful, data-driven study of NGC\,6752.

The collective works of Yong, Grundahl, and collaborators on NGC\,6752 are particularly useful in the sense that their data 
set includes sufficient information to piece together more-or-less the full picture of light element variations for more than 
20 red giants. Furthermore, the abundance of nitrogen is derived not from CN, but from NH,
whose measurement is independent of the carbon abundance. Sufficient information is present in this data set to accurately 
model, in detail, the stars in NGC\,6752 with very little in the way of assumptions regarding the abundances that are
significant for stellar evolution models \citep{Dotter2007,VandenBerg2012} as well as for synthetic spectra and
the associated bolometric corrections that are necessary to compare stellar evolution models to photometry.

The \emph{HST} photometry presented by \citet{Milone2013} comprises data reaching down the MS in 15
filters from $F225W$ in the the ultraviolet (UV) to $F160W$ in the near-infrared (near-IR); see their Table 1 for details. 
The multiple sequences revealed 
in these data exhibit a complex range of behaviors. Based on these observations, Milone et al.\ have identified
three stellar populations (labeled A, B, and C) that can be traced through different evolutionary phases in the 
CMDs. The RGB stars in these populations can also be matched with the spectroscopic targets of Yong et al.; 
this allows average abundances to be estimated for each population \citep[Table 2 of][]{Milone2013}.

The goal of the paper is to use the available abundance measurements to compute self-consistent stellar 
atmosphere and evolution models of stars at either end of the range of abundance variations (that is, for 
populations A and C) and then compare the \emph{HST} photometry with those models along the MS. 
The stellar atmospheres and synthetic spectra are computed with two independent codes. 
If the models are successful in tracing the observed sequences, then they constitute a powerful tool for 
interpreting the observed behavior in terms of the physical conditions found along the MS of NGC\,6752
and, by extension, other GCs for which observations of comparable quantity may be obtained.

The remainder of the paper is organised as follows: Section \ref{abund} lays out the abundances 
that will be used to construct models; full details are included as an Appendix. Section \ref{atm} describes the 
stellar atmosphere and spectrum synthesis codes and the models made by them. Section \ref{dsep} describes the 
stellar evolution models and isochrones. Section \ref{hst} presents detailed comparisons of the transformed 
isochrones with the \emph{HST} photometry. Section \ref{conclusions} summarises the important results and 
discusses future directions for this work.

\section{Elemental Abundances in NGC\,6752}\label{abund}
We have adopted the collection of abundances reported by \citet{Yong2003,Yong2005,Yong2008}. In particular, we
refer to the photometric identification of stellar populations A and C by \citet{Milone2013} and
their average abundances summarised in Table 2 of that paper. 
While \citet{Milone2013} have also identified an intermediate population (B), we have chosen not to include it in 
this study because our main goal is to discover how well stellar models are able to describe stars at the upper
and lower extremes of the light-element abundance distributions. If the models are able to reproduce the observed features in
populations A and C, then we are confident they will perform equally well when applied to population B.

In addition to the published abundances, we have added measurements of
carbon in 14 red giants (Yong et al., in preparation) in order to derive representative carbon abundances for populations 
A and C. Since the spectroscopic measurements do not include Ne, we have set [Ne/Fe]=+0.4 
in both populations. We have not included variations in elements heavier than copper because these elements are
under-abundant to begin with and show only slight variations (not more than 0.2 dex). These heavier elements are 
expected to have negligible influence on the properties of stellar evolution models in the H-R diagram and CMD. 
The results from Yong et al.\ (in preparation) indicate that the total C+N+O abundance is constant 
to within the measurement uncertainties ($<0.1$ dex) across all populations in NGC\,6752.

\begin{table}
  \centering
  \caption{Average Abundance Ratios from Spectroscopy\label{tab:ratio}}
  \begin{tabular}{@{}lrr@{}}
    \hline
            & \multicolumn{2}{c}{Population} \\
    Ratio   &    A     &    C     \\
    \hline
    $\mathrm{[Fe/H]} $ & $-1.65$  &  $-1.61$ \\
    $\mathrm{[C/Fe] }$ & $-0.25$  &  $-0.70$ \\
    $\mathrm{[N/Fe] }$ & $-0.11$  &  $+1.35$ \\
    $\mathrm{[O/Fe] }$ & $+0.65$  &  $+0.03$ \\
    $\mathrm{[Na/Fe]}$ & $-0.03$  &  $+0.61$ \\
    $\mathrm{[Mg/Fe]}$ & $+0.51$  &  $+0.40$ \\
    $\mathrm{[Al/Fe]}$ & $+0.28$  &  $+1.14$ \\
    $\mathrm{[Si/Fe]}$ & $+0.27$  &  $+0.35$ \\
    $\mathrm{[S/Fe] }$ & $+0.25$  &  $+0.25$ \\
    $\mathrm{[Ca/Fe]}$ & $+0.21$  &  $+0.27$ \\
    $\mathrm{[Ti/Fe]}$ & $+0.10$  &  $+0.15$ \\
    $\mathrm{[V/Fe] }$ & $-0.34$  &  $-0.25$ \\
    $\mathrm{[Mn/Fe]}$ & $-0.50$  &  $-0.45$ \\
    $\mathrm{[Co/Fe]}$ & $-0.03$  &  $-0.06$ \\
    $\mathrm{[Ni/Fe]}$ & $-0.06$  &  $-0.03$ \\
    $\mathrm{[Cu/Fe]}$ & $-0.66$  &  $-0.60$ \\
    \hline
  \end{tabular}
\end{table}

The average abundance ratios adopted for populations A and C are listed in Table \ref{tab:ratio}. A full listing of the 
abundances used in the paper, including number and mass fractions, is given in Appendix A.

The abundances reported by Yong et al.\ are based solely on RGB stars in NGC\,6752. It is therefore 
worthwhile to consider whether or not these abundances are appropriate for stars on the MS. Indeed, there is 
abundant evidence \citep{Smith2005A,Carretta2005,Smith2005B,Smith2006} that the CN-cycle operating in the H-burning shell, 
combined with deep mixing as GC stars climb the RGB, influences the surface abundances of carbon and nitrogen
while leaving oxygen essentially unchanged.

In the CN-cycle scenario the MS stars should have a higher carbon abundance, and a lower nitrogen abundance, than what is 
observed in the red giants while maintaining a constant sum. In the case of population A, carbon and nitrogen differ by $\sim0.5$ dex 
(see Table \ref{tab:abund}). In order to maximise the potential effect of the CN-cycle on population A while maintaining 
a constant sum, we consider a case in which nitrogen is depleted by 2 dex and carbon enhanced by 0.15 dex. We shall refer to this case 
as `ACN' hereafter. In the case of population C, for which C+N+O is already dominated by nitrogen, any changes due to the 
CN cycle should be very small because C+N+O will still be dominated by nitrogen, even after some of that nitrogen is converted back to 
carbon. We have tested this assertion and found that it does not produce any substantial effect in any of the models presented
in later sections; it will not be considered further.

Finally, we address the possibility of enhanced helium in population C. A spectroscopic study of horizontal branch (HB) stars 
in NGC\,6752 found no evidence for helium variation ($\mathrm{Y}=0.245\pm0.012$) but it was limited to stars on the red side 
of the HB, which are expected 
to retain the primordial helium abundance \citep{Villanova2009}. From careful consideration of colour differences in their full 
range of \emph{HST} photometry, \citet{Milone2013} estimated an increase in the helium mass fraction of $\dY= 0.03$. In order 
to quantify the influence of slightly enhanced helium at the level of $\dY=0.03$ we have computed additional models for 
population C with Y=0.28; we shall refer to this case as `$\CdY$'.

In summary, there are 4 sets of models that will be described and compared in the following sections: 
\begin{itemize}
\item case A is defined in Table \ref{tab:ratio} with a helium mass fraction $\mathrm{Y\approx0.252}$;
\item case C is also defined in Table \ref{tab:ratio} with $\mathrm{Y\approx0.252}$;
\item case ACN follows the pattern of case A with adjustments to carbon ($+0.15$ dex) and nitrogren ($-2$ dex) to account 
  for the CN-cycle in red giants while maintaining constant C+N; and
\item case $\CdY$ follows case C with an enhancement to helium such that $\dY=0.03$.
\end{itemize}

\section{Stellar Atmospheres Models}\label{atm}
The abundance profiles described in Section \ref{abund} and listed in Appendix \ref{app} were used to construct model atmosphere
structures and synthetic spectra with two different codes.

  \subsection{ATLAS and SYNTHE}
  The ATLAS models for cases A and C were computed with the ATLAS12 model atmosphere code, part of the Kurucz lineage 
  of atmosphere routines \citep{Kurucz1970,Kurucz1993}, ported to Linux by \citet{Sbordone2004}. The grid of models 
  covers log(g)= 2, 3, 4, and 5 and $\Teff$= 3,500, 4,000, 4,500, 5,000, 5,750, and 6,500K. ATLAS12 employs the opacity 
  sampling technique to construct model atmospheres.  These are plane-parallel and assume local thermodynamic 
  equilibrium (LTE).  A microturbulent velocity of $2~\kms$ was adopted for all of the models. Synthetic spectra were 
  computed with SYNTHE \citep{Kurucz1981} at a resolution of R=500,000 from 1,000 to 30,000\AA. The latest atomic line 
  lists (kindly provided by R.\ Kurucz) were used, as were molecular line lists for H$_2$O, TiO, FeH, CrH, CaH, C$_2$, 
  CN, CH, NH, SiO, SiH, OH, MgH, CO, and H$_2$.\footnote{The line lists are currently undergoing a major update. Quantitative
  results based on these models, particularly in the UV, are subject to change.}

  \begin{figure*}
    \includegraphics[width=160mm]{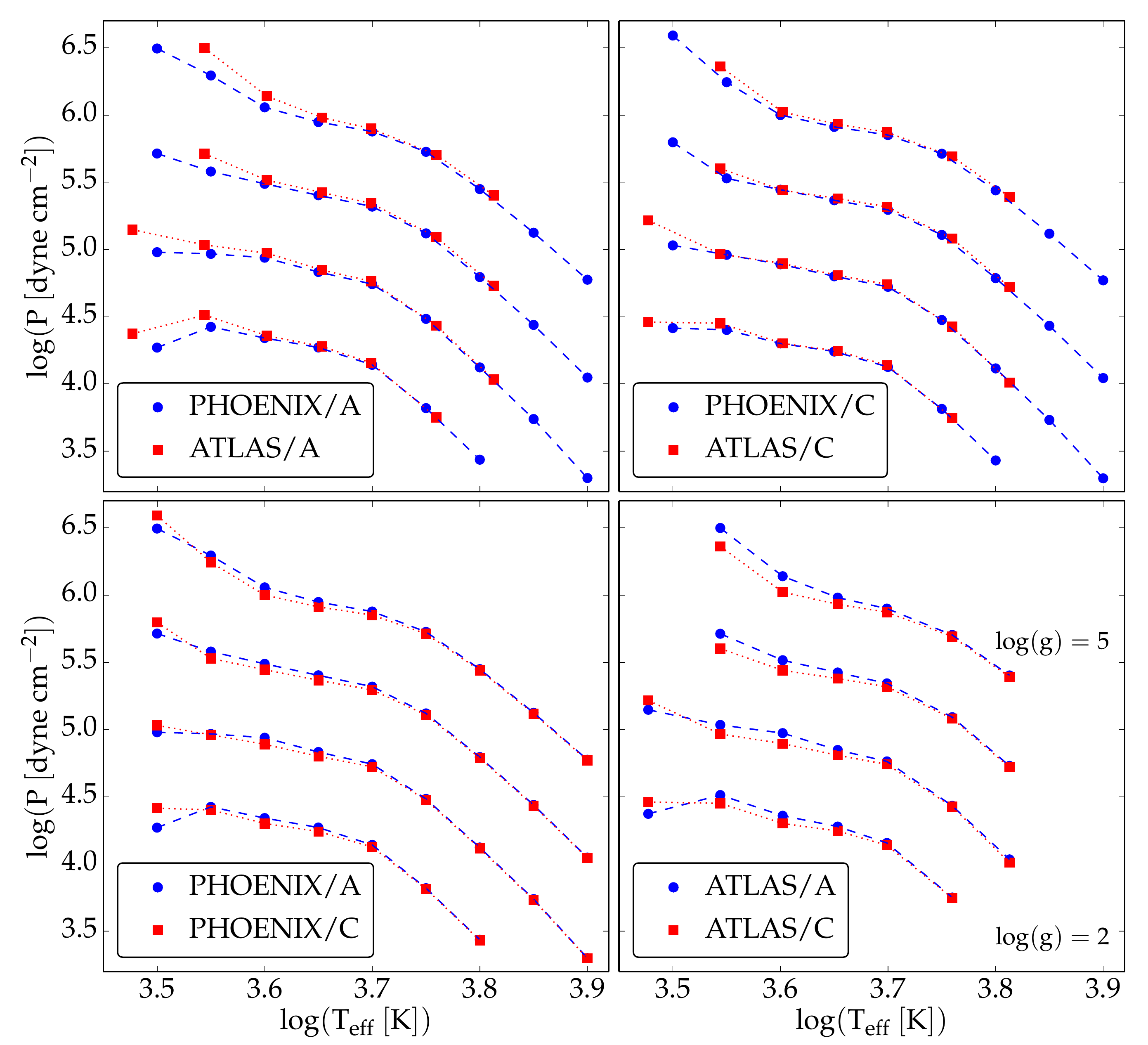}
    \caption{The log of pressure (cgs units) at T=$\Teff$ from \phx\ and ATLAS model atmospheres for cases A and C. The top left 
      and right panels compare \phx\ and ATLAS for case A and C, respectively. The bottom left and right panels compare 
      cases A and C for \phx\ and ATLAS, respectively. All panels show models for log(g)=2,3,4,5 from bottom to top; 
      log(g)=2 and log(g)=5 are labeled in the bottom right panel. All panels have the same dimensions.\label{fig:atm}}
  \end{figure*}

  \subsection{PHOENIX}
  A grid of model atmospheres and synthetic spectra computed using the \phx\ code version discussed by
  \citet{Hauschildt1999A,Hauschildt1999B} with updates described by \citet{Ferguson2005}.  The individual models are 
  plane-parallel, assume LTE, and have a microturbulent velocity of $2~\kms$. The resolution of the synthetic spectra 
  differs by region; the one relevant to this study is 0.2\AA\ between 1,000 and 20,000\AA.
  Grids of atmospheres and synthetic spectra were generated for cases A and C with effective temperatures from 
  $\log(\Teff) = 3.50$ ($\sim3,100$ K) to 3.90 ($\sim8,000$ K), in   steps of 0.05 dex and values of log(g) from 2.0 
  to 5.5 in steps of 0.5 dex. Additional models were computed for cases ACN and $\CdY$ for the full range of 
  $\log(\Teff)$ given above, at log(g)=5, to provide comparisons with cases A and C for conditions appropriate for 
  stars on and near the MS.

  \begin{figure*}
    \includegraphics[width=160mm]{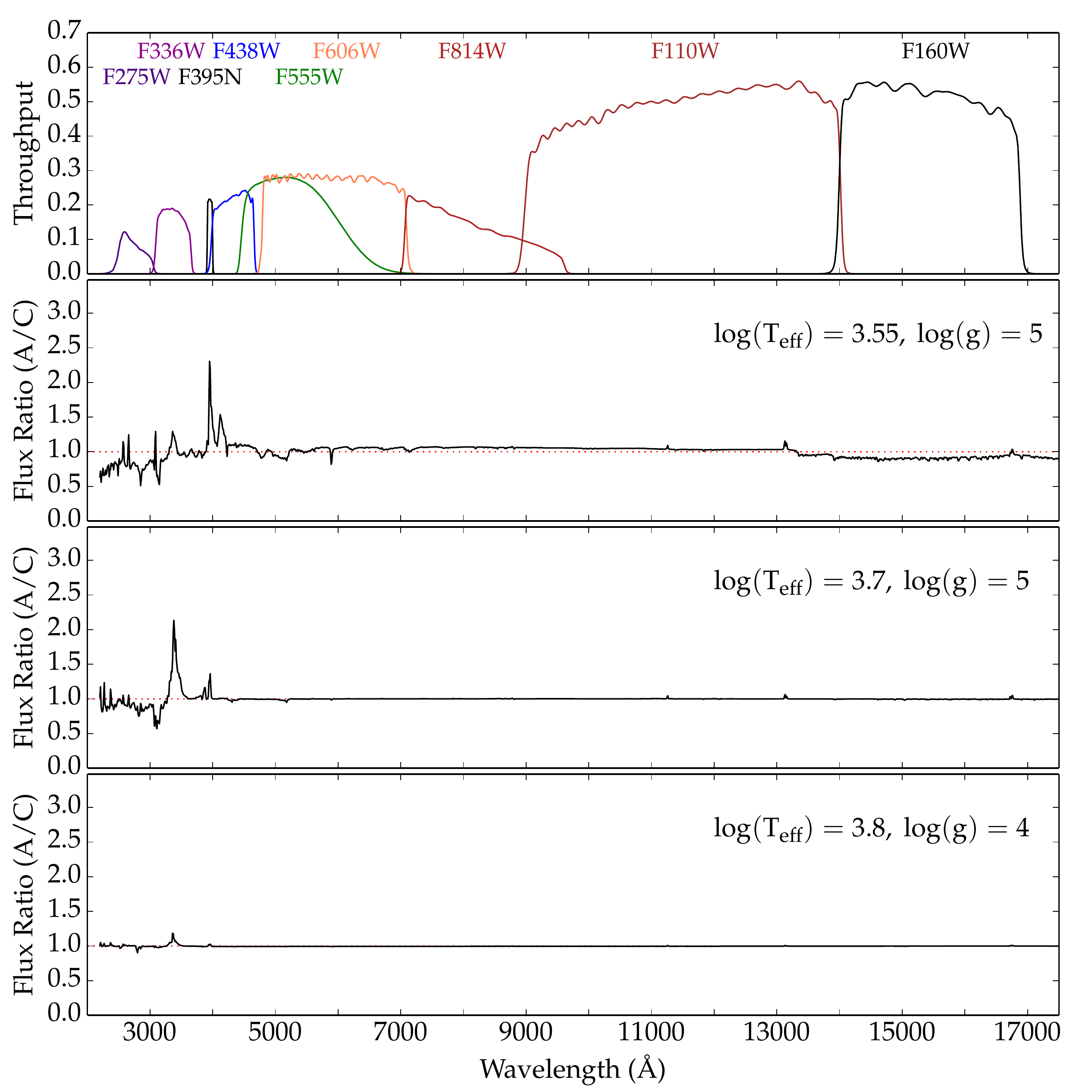}
    \caption{\emph{HST} WFC3 filter throughputs (top) and flux ratios, in the sense of case A over case C, for three 
      pairs of \phx\ spectra. The wavelength-dependent flux has been smoothed for clarity (see text for details). The 
      coolest star shows substantial differences for wavelengths shorter than about 5,000\AA\ and longer than
      about 13,000\AA. The other stars show differences exclusively for wavelengths shorter than 4,000\AA.
      \label{fig:phxflux}}
  \end{figure*}

  \begin{figure*}
    \includegraphics[width=160mm]{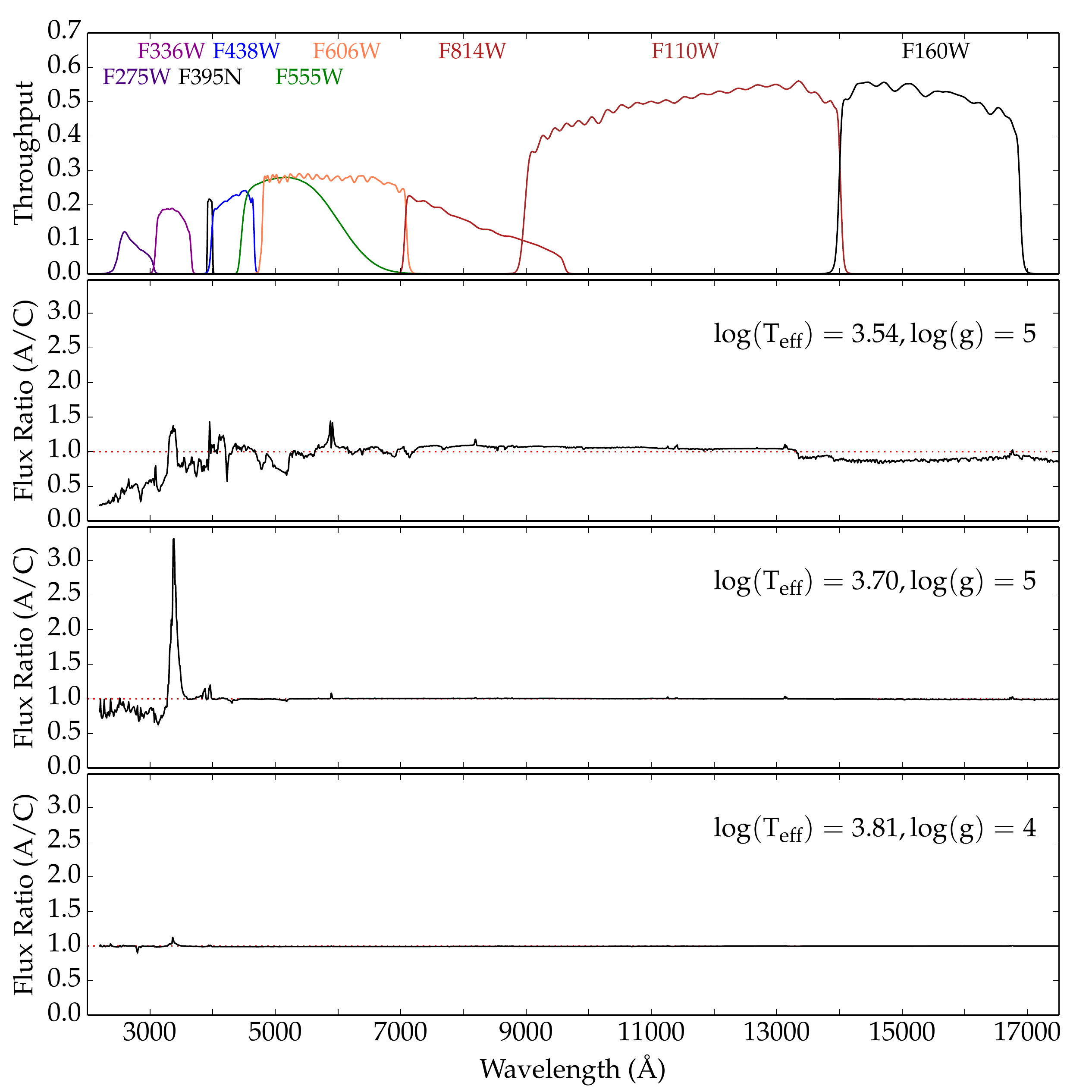}
    \caption{\emph{HST} WFC3 filter throughputs (top) and flux ratios, in the sense of case A over case C, for three 
      pairs of ATLAS/SYNTHE models. The wavelength-dependent flux has been smoothed for clarity. The dimensions of 
      each panel are identical to those in Figure \ref{fig:phxflux}. The results are mostly simlar to the \phx\ models 
      shown in Figure \ref{fig:phxflux} but the ATLAS/SYNTHE models show stronger differences in the optical spectrum 
      for the coolest star.\label{fig:kurflux}}
  \end{figure*}

  \begin{figure*}
    \includegraphics[width=160mm]{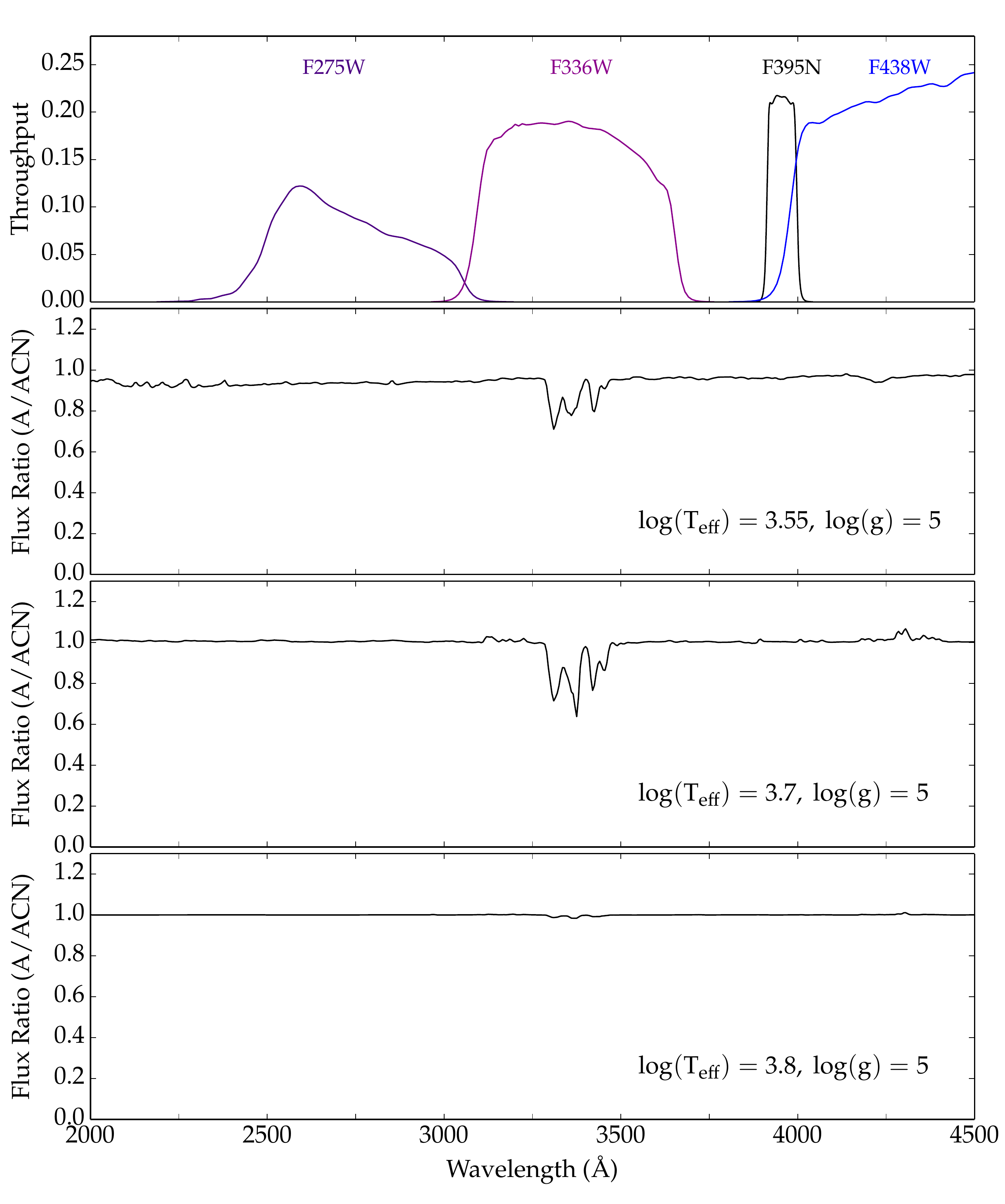}
    \caption{\emph{HST} WFC3 filter throughputs (top) and flux ratios, in the sense of case A to case ACN, 
      for three \phx\ models. The wavelength-dependent flux has been smoothed for clarity. Flux differences
      due to CN-cycling on the RGB appear mostly in the $F336W$ filter. Note that the x- and y-axes show
      smaller regions than in Figures \ref{fig:phxflux} and \ref{fig:kurflux}.\label{fig:CNflux}}
  \end{figure*}

  \begin{figure*}
    \includegraphics[width=160mm]{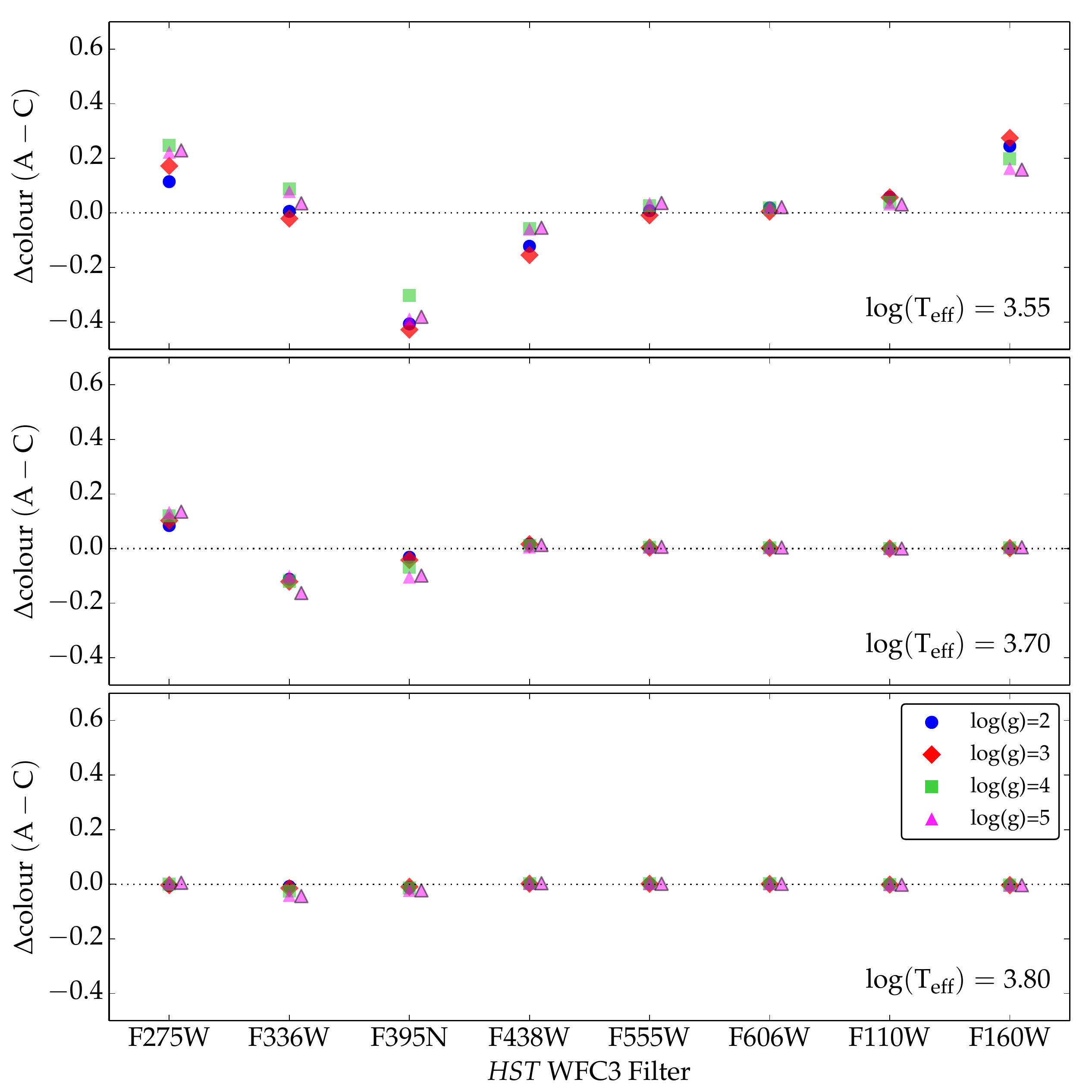}
    \caption{Colour difference, defined with respect to $F814W$, between cases A and C for the \phx\
      synthetic spectra shown in Figure \ref{fig:phxflux}. Colour difference between cases ACN 
      and C, for log(g)=5 only, are shown offset slightly to the right and with a black outline.
      Case ACN is only relevant in F336W as should be evident from Figure \ref{fig:CNflux}.
      \label{fig:phxcolor}}
  \end{figure*}

  \begin{figure*}
    \includegraphics[width=160mm]{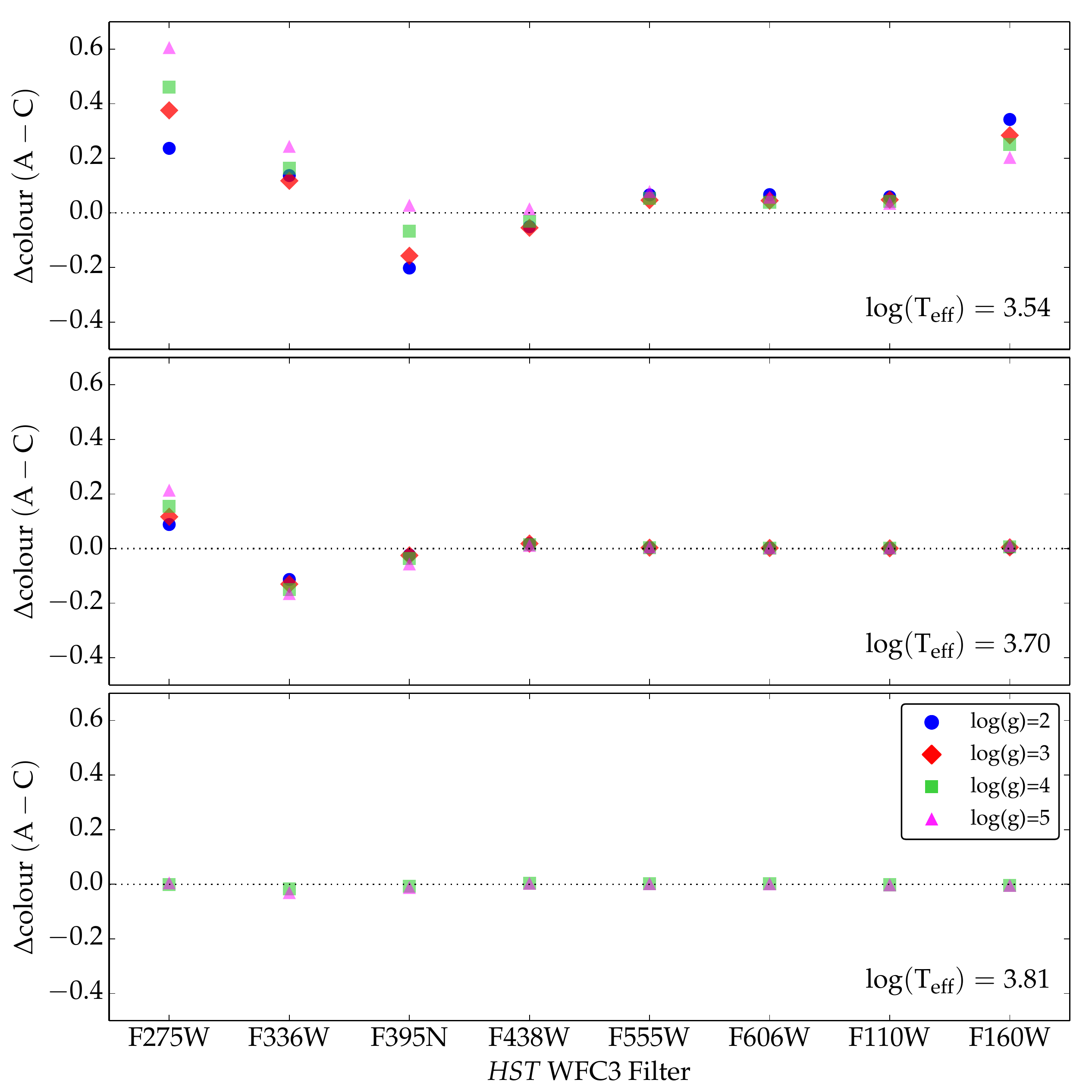}
    \caption{Colour difference, defined with respect to $F814W$, between cases A and C for the SYNTHE 
      synthetic spectra shown in Figure \ref{fig:kurflux}. The dimensions of each panel are the
      same as in Figure \ref{fig:phxcolor}.\label{fig:kurcolor}}
  \end{figure*}

  \subsection{Comparisons}
  The model atmosphere structures differ only slightly between cases A and C, with departure growing as 
  temperature decreases. Figure \ref{fig:atm} shows the photosphere pressure (extracted at T=$\Teff$) from both \phx\ 
  and ATLAS models for a range of log(g) and cases A and C. The pressure is used as the surface boundary condition 
  in the stellar evolution models, see Section \ref{dsep}. The upper panels of Figure \ref{fig:atm} compare \phx\ 
  and ATLAS pressures for case A (left) and case C (right); the bottom panels compare cases A and C from \phx\ 
  (left) and ATLAS (right). The upper panels indicate that \phx\ and ATLAS yield very similar pressures except 
  for the lowest temperature point in the ATLAS grid. The lower panels indicate that cases A and C give essentially
  the same pressures above 5,000K (log($\Teff$[K])=3.7). Between 3,500 and 5,000K case A has consistenty higher 
  pressures; below 3500K the situation reverses and case C has the higher pressure.
  \comment{The difference in pressure for the coolest stars is due to differences in the molecular
    composition which, in turn, is driven by chemical equilibrium as set by the respective equations
    of state. Difference in composition leads to difference in opacity and, ultimately, to difference
    in the atmosphere structure.}

  Figure \ref{fig:phxflux} shows the throughput of several \emph{HST} WFC3 filters, in the top panel, 
  followed by the ratio of the fluxes (measured in $\mathrm{erg\,cm^{-2}\,s^{-1}}$\AA$^{-1}$) from \phx\ 
  synthetic spectra for cases A and C at fixed temperatures and gravities. The flux is plotted every 
  10\AA\ and has been smoothed with a 3\AA\ Gaussian filter for improved clarity.
  The temperatures and gravities are representative of points along the MS and chosen to
  correspond, as closely as possible, to models in the ATLAS/SYNTHE grid. Figure \ref{fig:kurflux} 
  is the equivalent of Figure \ref{fig:phxflux} for the ATLAS/SYNTHE grid; the SYNTHE spectra were
  smoothed in the same way as the \phx\ spectra.

  Figures \ref{fig:phxflux} and \ref{fig:kurflux} both show, for the coolest temperature, that there
  are significant differences in the fluxes between cases A and C below 5,000 \AA\ and above 13,000 
  \AA. For the hotter temperatures the differences are restricted to wavelengths shorter than 4,000 \AA.
  The difference in the coolest models' flux ratio in the near-infrared is due to suppressed H$_2$O 
  absorption because of the reduction of oxygen in case C compared to case A. SiO absorption in the UV is 
  an important factor in the flux ratio blueward of 3,000\AA. NH is responsible for the peak near 3,300\AA.

  The contribution of CN in the coolest models shown in Figures \ref{fig:phxflux} and \ref{fig:kurflux} 
  is negligible because the majority of carbon is locked up in CO; this is true in cases A and C because oxygen 
  outnumbers carbon in both cases. In this way the MS differs from the RGB, where CN
  variations have been observed for many years \citep[][]{FreemanNorris1981}.

  Figure \ref{fig:CNflux} shows the flux ratio of cases A and ACN, similar to Figures 
  \ref{fig:phxflux} and \ref{fig:kurflux} and with the same smoothing applied, though the 
  differences are much more restricted in this case. The species responsible for the only significant 
  variations in Figure \ref{fig:CNflux} is NH. We were able to verify this by computing additional 
  \phx\ models without NH for the same abundances and physical conditions: the flux ratio for these 
  models is flat (ratio=1) through the same wavelength region. Whereas the differences between cases 
  A and C will alter synthetic photometry in UV and IR filters, the difference between cases A and 
  ACN will only \comment{modestly} influence the $F336W$ filter. 
  \comment{ For the MS stars considered here, the difference between cases A and ACN 
    in the $F336W$ filter amounts to, at most, 0.1 mag at about 4,200K. The difference decreases to 
    less than 0.02 mag above 5,500K and below 3,000K. In both $F275W$ and $F438W$ the same models
    never differ by more than 0.02 mag.}

  The flux ratios shown in Figures \ref{fig:phxflux} and \ref{fig:kurflux} will have a measurable influence 
  on the bolometric corrections, hence colours, derived from the synthetic fluxes. Figures \ref{fig:phxcolor} 
  and \ref{fig:kurcolor} show the colour difference between cases A and C for a selection of 9 \emph{HST} 
  WFC3 filters. These figures are modeled after, and should be compared with, Figures 17 and 18 of \citet{Milone2013} where
  they are used to demonstrate the separation between sequence in different colours.
  Figure \ref{fig:phxcolor} also shows, offset to the right, the colour difference between cases ACN 
  and C for log(g)=5 only. The temperatures shown are the same as in Figure \ref{fig:phxcolor}
  and \ref{fig:kurcolor}, respectively, for a range of surface gravities. In Figures \ref{fig:phxcolor} and \ref{fig:kurcolor},
  $\Delta$colour refers to the colour derived from the case A synthetic spectrum minus the colour derived from the 
  case C synthetic spectrum for the same physical parameters.  The colours are defined with respect to $F814W$ as 
  $\mathrm{M_{Filter}-M_{F814W}}$.\footnote{We acknowledge that what is shown here from the models is really a difference 
  in magnitude--not colour--for a given filter, taken at a fixed $F814W$ magnitude. However, we have chosen to 
  present it as a colour difference in accordance with \citet{Milone2013}. The main reason is that this measurement
  would be done in terms of colour at fixed magnitude in an observed CMD.}

  Both Figures \ref{fig:phxcolor} and \ref{fig:kurcolor} indicate that the warmer spectra (the 2 bottom panels)
  show colour differences measured in hundredths of magnitudes for the optical and infrared filters; the only
  significant differences for these spectra are in the UV. For the coolest spectra the colour differences are more 
  substantial for all filters except $F555W$ and $F606W$. The effect of accounting for CN cycle modifications to the 
  surface abundances results in non-negligible differences only to $F336W$, consistent with Figure \ref{fig:CNflux}.

  \begin{figure*}
    \includegraphics[width=140mm]{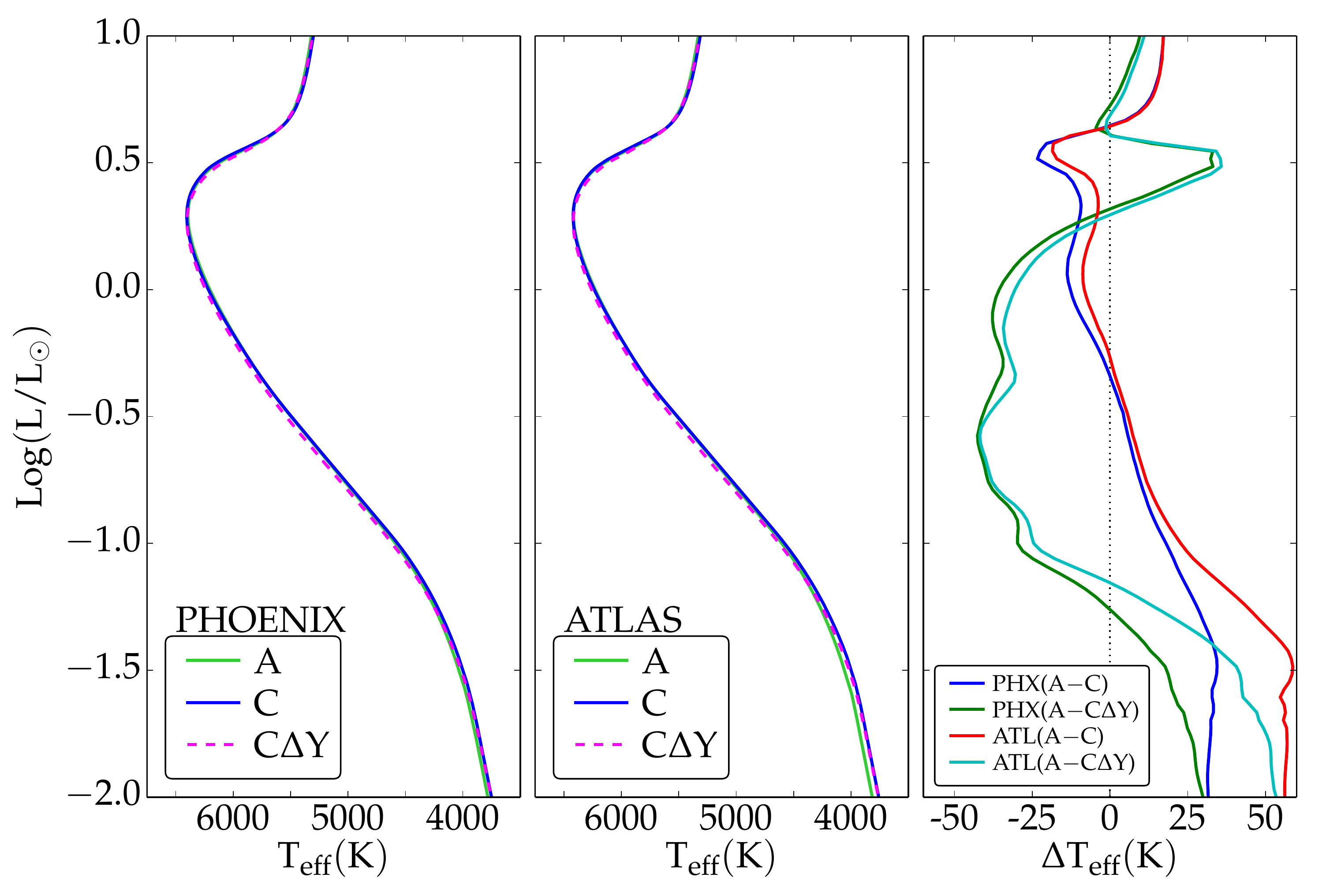}
    \caption{\isoage\ Gyr isochrones computed for cases A, C, and $\CdY$ with atmosphere boundary conditions from 
      \phx\ (left) and ATLAS (center). The dimensions of both panels are the same. 
      \comment{The right panel shows the temperature difference between cases A and C and A and C$\dY$ at fixed luminosity.
      Here \phx\ is abbreviated PHX and ATLAS by ATL.}
      Noticeable differences between A and C appear only as $\Teff$ falls below 5,000K and molecules become important 
      in the atmsophere. $\CdY$ shows the characteristic appearance of stellar models with slightly enhanced helium: 
      marginally hotter along the MS with a steeper slope through the subgiant branch.\label{fig:iso}}
  \end{figure*}

  \begin{figure*}
    \includegraphics[width=180mm]{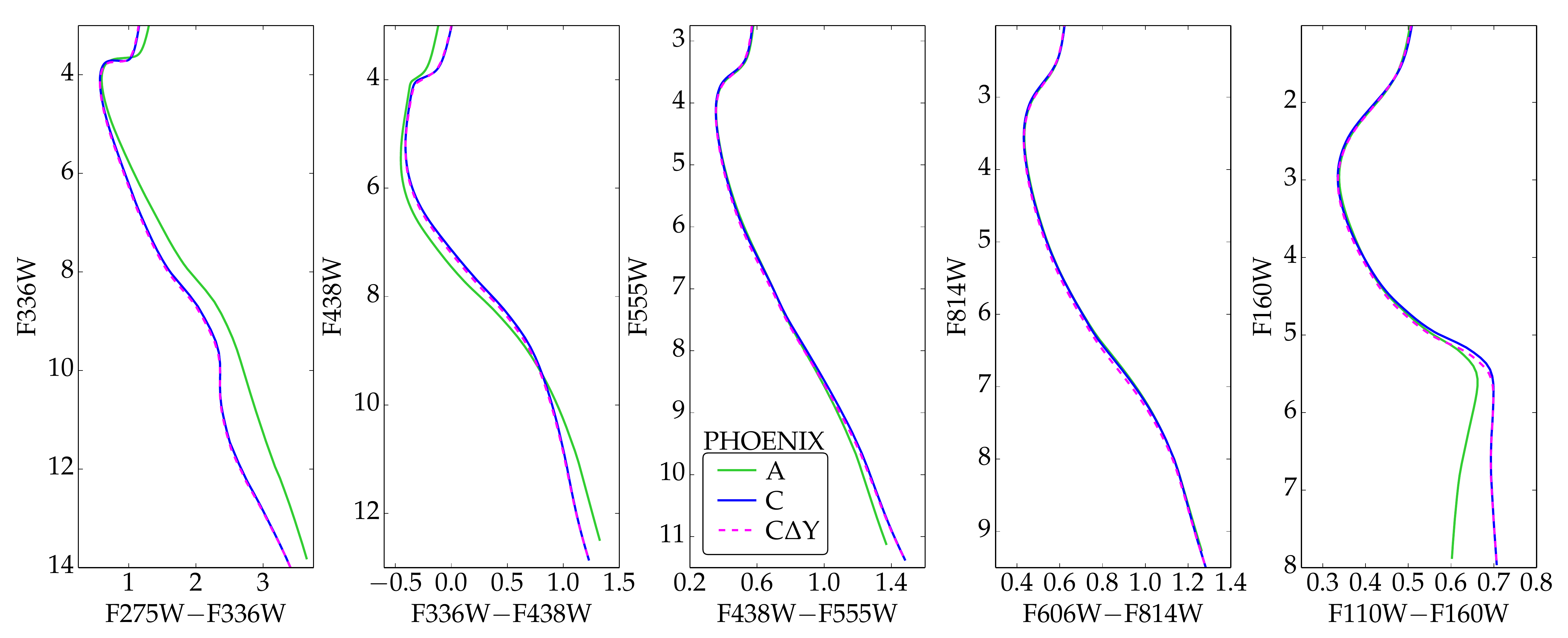}
    \caption{The \isoage\ Gyr \phx\ isochrones shown in Figure \ref{fig:iso} with \emph{HST} WFC3 bolometric corrections 
      from the \phx\ synthetic spectra. The key in the middle panel applies to all.\label{fig:isophxWFC3}}
  \end{figure*}

  \begin{figure*}
    \includegraphics[width=180mm]{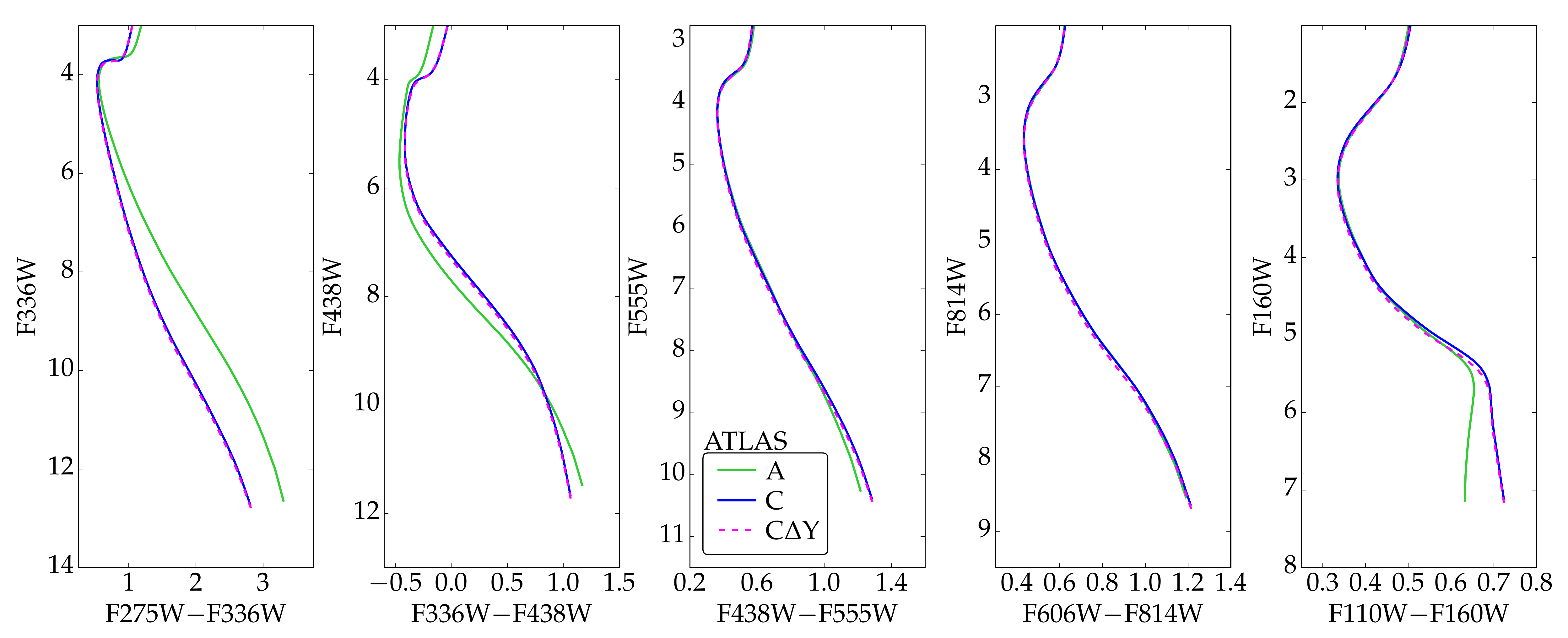}
    \caption{Equivalent to Figure \ref{fig:isophxWFC3} for the \isoage\ Gyr ATLAS isochrones with bolometric corrections 
      from the SYNTHE spectra. The dimensions of each panel are the same as the corresponding panel in Figure 
      \ref{fig:isophxWFC3}; the models end at brighter magnitudes than those in Figure \ref{fig:isophxWFC3} because the 
      SYNTHE spectra are limited to $\Teff \geq 3,500$K. The key in the middle panel applies to all.\label{fig:isokurWFC3}}
  \end{figure*}

\section{Stellar Evolution Models and Isochrones}\label{dsep}

Stellar evolution models were computed with the Dartmouth Stellar Evolution Program (DSEP), which has been 
configured to self-consistently account for specific chemical abundance patterns in the nuclear reactions, 
equation of state, low- and high-temperature opacities, and surface boundary condition \citep{Dotter2007}. 
The model atmospheres described in Section \ref{atm} were used to derive the surface boundary condition, the 
photosphere pressure, at the point in the atmosphere where T=$\Teff$ for each composition and from each 
atmosphere code (see Figure \ref{fig:atm} and related text).

All details of the code remain as stated by \citet{Dotter2007B,Dotter2008} and we list here only those elements 
of the code most relevant to this study: DSEP uses the FreeEOS equation of 
state,\footnote{http://freeeos.sourceforge.net} high-temperature opacities from OPAL \citep{OPAL1996}, and
low-temperature opacities from \phx\ \citep{Ferguson2005}. The opacities were computed for the abundance
patterns described in Section \ref{abund} (see Appendix \ref{app} for full details).

The Rosseland mean low-temperature opacities differ between cases A and C by no more than 2\% above 5,500K; since the stellar 
evolution models cooler than 5,500K have substantial convective envelopes, and use model atmosphere boundary conditions,
 the influence of the low-temperature opacities is minimal. The same goes for the high-temperature opacities, where the 
largest---though still quite modest---difference occurs around $10^6$K.

Stellar evolution tracks with masses from 0.2 to 0.8 $\Msun$ were computed starting from the fully-convective 
pre-MS and ending at a point suitable for the construction of isochrones with ages appropriate for
GCs, namely 10-14 Gyr.
The evolutionary tracks were transformed into isochrones in the same manner as those of the Dartmouth Stellar 
Evolution Database \citep{Dotter2007B,Dotter2008} and then converted to observable magnitudes by interpolation 
in bolometric correction tables that were derived from the synthetic spectra described in Section \ref{atm}. 

Figure \ref{fig:iso} shows \isoage\ Gyr isochrones computed using the surface boundary condition from \phx\ 
models (left) and ATLAS models (center) for cases A, C, and $\CdY$ (the last of which uses the same
boundary condition as case C). \comment{The right panel shows the temperature difference at fixed luminosity between different cases.
The plot shows similar behavior in both panels: case A and case C overlap almost perfectly 
from the base of the RGB through the MS turnoff and down the MS until $\Teff$ 
reaches about 4,500K. Below 4,500K, corresponding to $\log(L/\Lsun) \sim -1$, the onset of molecules in the atmospheres begins
to push cases A and C apart with A remaining slightly hotter---but never by more than about 50K---than C down to the extent of 
the models shown ($0.2\Msun$). The \phx\ models show a smaller separation than the ATLAS models; this can be traced back to 
the difference in surface pressures seen in Figure \ref{fig:atm}. The $\CdY$ isochrone is consistently hotter than the C 
isochrone below the MSTO and above the molecular regime, which is typical behavior for slightly helium-enhanced stellar evolution 
models, but the difference is only at the level of 50K.}

  \begin{figure*}
    \includegraphics[width=180mm]{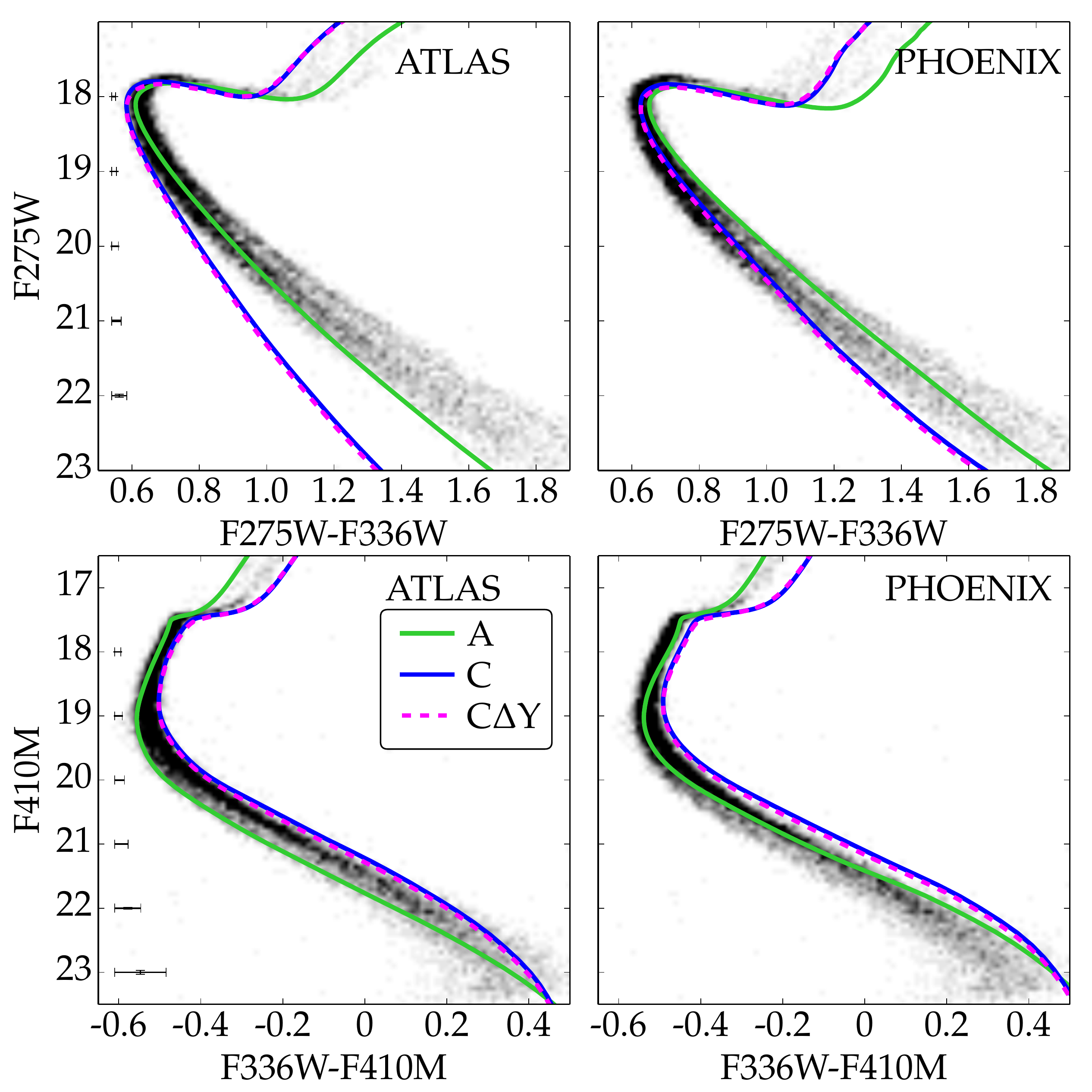}
    \caption{The top row shows the $F275W-F336W$ CMD compared with ATLAS isochrones (left) and \phx\ isochrones
      (right). The bottom row shows the analogous panels for the $F336W-F410M$ CMD. Typical photometric errors
      are shown on the left edge of the lefthand panel in each row. \label{fig:Hess}}
  \end{figure*}

The similarity evident in Figure \ref{fig:iso} indicates that the observed differences in various colour combinations
are due to the influence of abundance variations on the stellar spectra with only a modest
contribution from the difference in $\Teff$ on the lower MS. Figure \ref{fig:iso} reiterates the point
already made by \citet{Pietrinferni2009} that the influence of light-element variations on isochrone ages is negligible
so long as the total C+N+O content is constant among the different stellar populations.

  \begin{figure*}
    \includegraphics[width=180mm]{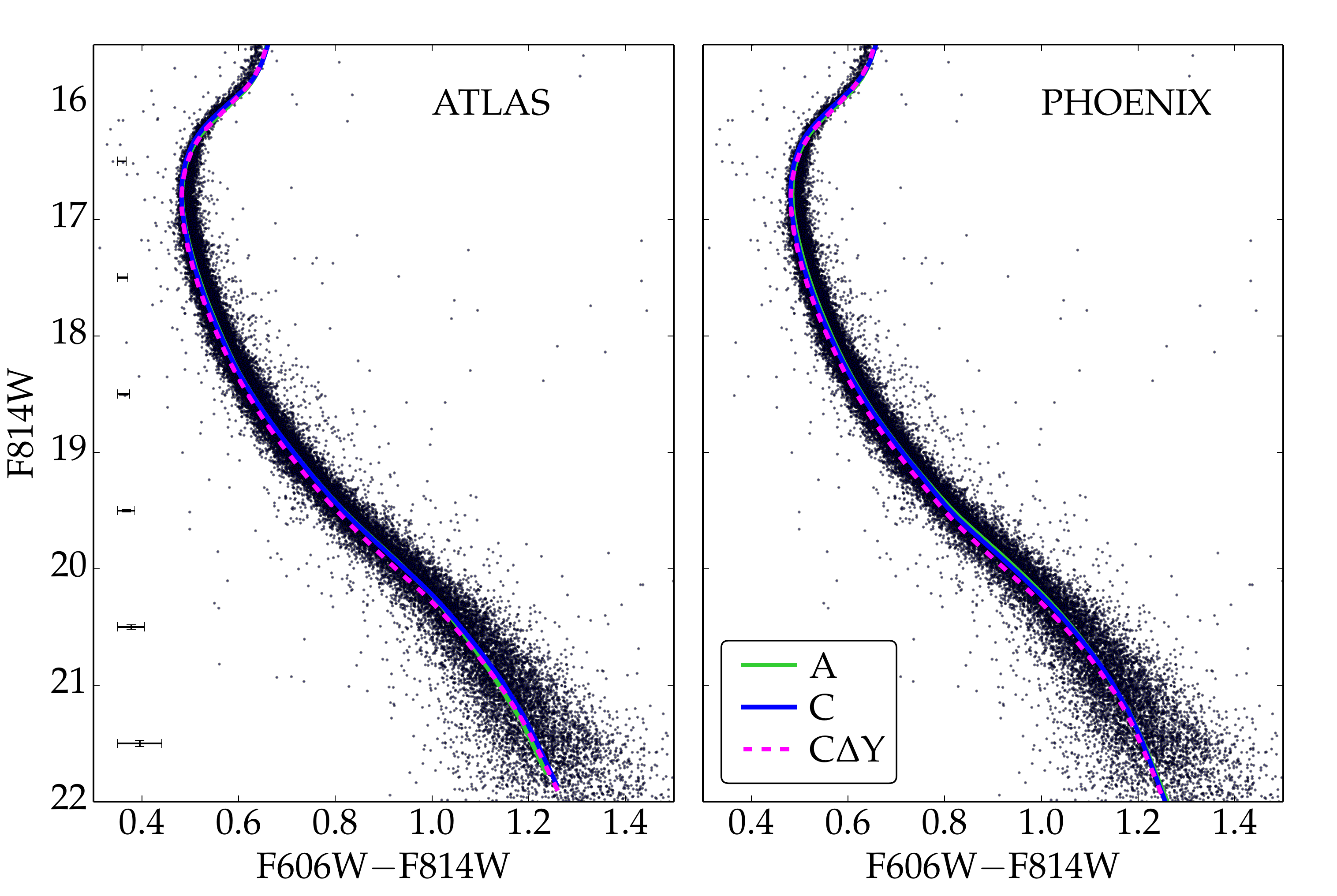}
    \caption{Isochrones from populations A and C compared to the \emph{HST} ACS $F606W-F814W$ CMD 
      \citep{Sarajedini2007,Anderson2008}. Typical photometric errors are shown on the left edge 
      of the lefthand panel. Both sets of models show no obvious difference except 
      that the helium-enhanced models are very slightly hotter along the MS.\label{fig:ACS}}
  \end{figure*}

Isochrones transformed into 5 broadband CMDs from the \emph{HST} WFC3 (Vegamag) photometric system are presented in Figure 
\ref{fig:isophxWFC3} for bolometric corrections from \phx\ models and Figure \ref{fig:isokurWFC3} for ATLAS/SYNTHE 
models. Qualitatively, each CMD shows the same behavior from the \phx\ and ATLAS/SYNTHE models. In the UV and blue
filters there are substantial differences between cases A and C; in the optical CMDs the differences are minimal; and
the near-IR cases A and C separate below the `knee' on the lower MS. More detailed discussion of some of these features
will be given in Section \ref{hst} where the isochrones are compared with the \emph{HST} photometry of \citet{Milone2013}.

  \begin{figure*}
    \includegraphics[width=180mm]{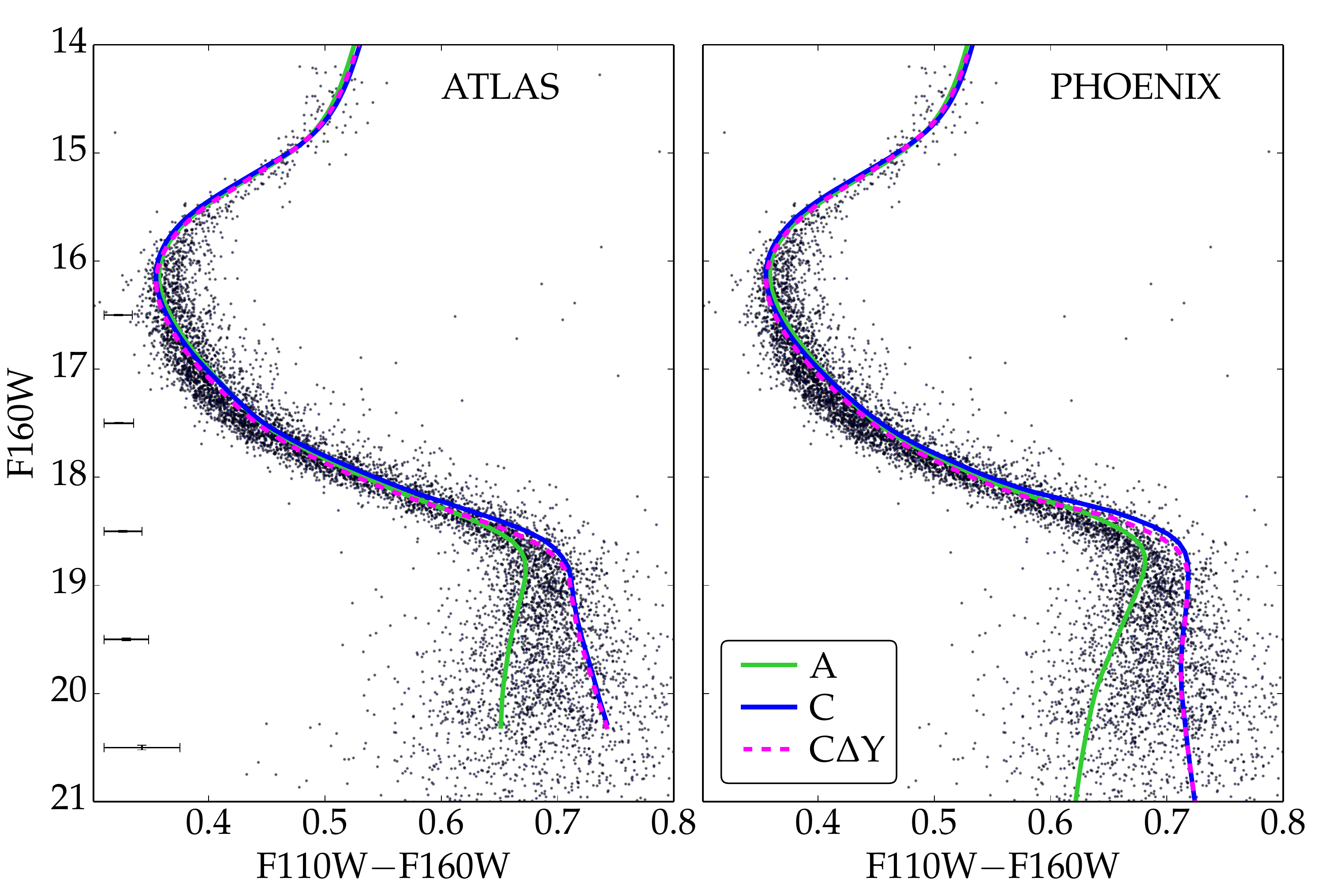}
    \caption{Isochrones for populations A and C compared to the \emph{HST} $F110W-F160W$ CMD. 
      Typical photometric errors are shown along the left edge of the lefthand panel.
      The separation of populations A and C is evident in both diagrams, with both sets of models 
      showing population A bluer than C below the MS knee.\label{fig:CMDIR}}
  \end{figure*}

\section{Comparisons with \emph{HST} photometry}\label{hst}
This section will demonstrate the extent to which the isochrones presented in Section \ref{dsep} 
are able to match the \emph{HST} photometry of NGC\,6752 by \citet{Milone2013} along the MS. Throughout we have 
adopted an isochrone age of \isoage\ Gyr for both populations, a true distance modulus of $(m-M)_0 = 13.13$ 
\citep[][specifically the 2010 revision]{Harris1996}, and a reddening value of $E(B-V)=0.0485$ \citep{SF11}. 
We have used the \citet{CCM} extinction curve with the adopted reddening value and $R_V=3.1$, corresponding to
 $A_V=0.15$, to derive the extinction along with the bolometric correction in each bandpass for each synthetic spectrum.

\subsection{UV and Blue CMDs}

The comparisons begin with $F275W-F336W$ (upper panels) and $F336W-F410M$ (lower panels) in Figure \ref{fig:Hess}. In
both rows of Figure \ref{fig:Hess} the left panel shows the ATLAS isochrones with SYNTHE bolometric corrections and the 
right panels show the \phx\ isochrones. The data are shown as Hess diagrams in order to better display the split sequences.

In the $F275W-F336W$ CMD both sets of isochrones provide a reasonable description of data from the 
turnoff to about 2 (ATLAS/SYNTHE) or 3 (\phx) magnitudes fainter before they diverge blueward of the data. The separation 
is more dramatic in the ATLAS/SYNTHE case. Comparison of the two sets of isochrones with the Hess diagram shows that the
separation between populations A and C increases slowly from $\Delta(F275W-F336W)\sim0$ at the turnoff to 
$\Delta(F275W-F336W)$=0.2-0.3 at the limit of the diagram ($F275W=23$). 
\comment{While the isochrones do not match the data in an absolute sense, the relative separation in $F275W-F336W$ at fixed 
magnitude is, at least qualitatively, reproduced by both sets of models.}

In the $F336W-F410M$ CMD the isochrones stay within the envelope of the data points for almost the full extent of the CMD 
shown in Figure \ref{fig:Hess}. The separation between the populations within 2 mags of the turnoff is  reproduced 
by the models. In this case the ATLAS/SYNTHE models give an accurate representation of the data from the base of the
RGB to the lower limit shown at $F410M=23.5$. 
\comment{Both sets of models predict that the sequences should cross at $F410M\sim23$ but, unfortunately, the 
photometric errors are too large at this point, and the data runs out just below, to verify the model prediction.}

There are a variety of reasons why cool star models may be discrepant with the data in UV and blue spectral regions.
These include missing opacity from atoms and molecules, neglect of non-LTE effects, and phenomenological treatment of 
inherently 3D physics in stellar atmosphere (e.g., convection and microturbulence).
Suffice it to say that, while the models leave room for improvement in an absolute sense, both the general trends and also the 
relative difference between cases A and C are consistent with the observations.

\subsection{Optical and Near-IR CMDs}
\citet{Sbordone2011} pointed out that the influence of abundance variations on synthetic spectra and photometry are
significant in the UV and blue but leave the optical largely untouched. Figures \ref{fig:isophxWFC3} and \ref{fig:isokurWFC3}
suggest that the most prominent CMD in recent \emph{HST} GC age studies, \emph{HST} ACS $F606W-F814W$ 
\citep{Sarajedini2007,Anderson2008}, is insensitive to light-element abundance variations provided that C+N+O remains
constant.

Figure \ref{fig:ACS} compares the ACS photometry \citep{Anderson2008} with the isochrones in the $F606W-F814W$ CMD.
It should be clear from this comparison that there is essentially no difference between populations A and C in this
CMD and that, even with a small $\dY$ between populations A and C, this is perhaps the safest CMD in which to perform 
GC age analyses because it is the least sensitive to the photometric manifestation of light-element variations.

The synthetic spectra by \citet{Sbordone2011} extended only to 10,000\AA\ and, therefore, the near-IR filters 
were excluded from their analysis. Meanwhile observations of NGC\,2808 \citep{Milone2012} and M\,4 \citep{Milone2014}
show that the MS of both clusters fans out below the `knee' in the \emph{HST} WFC3/IR $F110W-F160W$
CMD. The same feature has also been observed in 47\,Tuc by \citet{Kalirai2012}. The broadening of the lower MS 
is attributed to variation in absorption by water, primarily in the $F160W$ filter 
\citep{Milone2012,Milone2014}. This conclusion is fully supported by the material presented in Section \ref{atm}.

We plot the isochrones on top of the \emph{HST} WFC3/IR CMD in Figure \ref{fig:CMDIR}. The CMD is based on the
same data as presented by \citet{Milone2013} but, 
\comment{unlike in \citet{Milone2013} where only stars found in all bandpasses were shown, we have included 
all stars that were measured in the near-IR observations. }
The models clearly 
indicate that the sequences should split just above the knee. Population A should have bluer $F110W-F160W$ 
colours than population C below the knee. This is consistent with the flux ratios of the coolest models shown in 
Figures \ref{fig:phxflux} and \ref{fig:kurflux}. 
  
  \begin{figure*}
    \includegraphics[width=180mm]{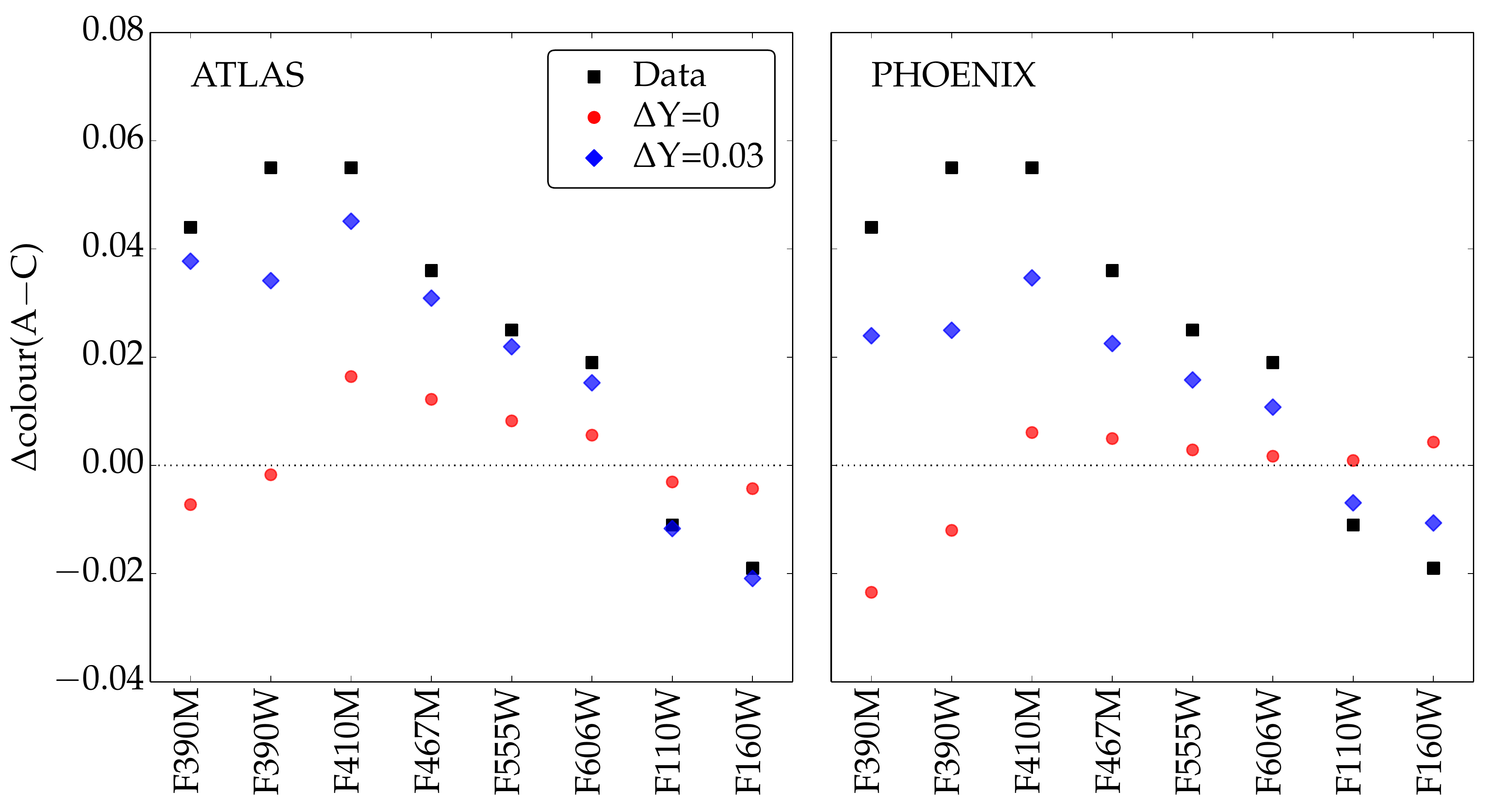}
    \caption{The colour differences between cases A and C in a variety of filters at $F814W=18.5$ for ATLAS/SYNTHE 
      isochrones (left) and \phx\ isochrones (right). These are compared with the equivalent measurement of populations 
      A and C by \citet{Milone2013}. The plot demonstrates the influence of enhanced helium on the relative colours
      of the two populations.\label{fig:MSY}}
  \end{figure*}

The comparison between models and data in Figure \ref{fig:CMDIR} suggests that the near-IR CMD can provide an estimate
of oxygen-variation within a GC such as NGC\,6752 provided that self-consistent models for the appropriate 
metallicity, and with a range of oxygen, are available. Furthermore, a proper accounting of the photometric errors in the
data is required to distinguish between a spread in the MS due to abundance variation or increasing photometric error.
The models show an obvious trade-off between the location at which the spread is measured and the sensitivity of the
near-IR colour. In the case of NGC\,6752 shown here, the spread increases from nearly zero at the knee to almost 0.1
in colour about 2 magnitudes below the knee. This technique may prove a useful first step in studying oxygen variation
in GCs which have, for example, few bright red giants or significiant extinction: either of which would complicate
a proper spectroscopy analysis. This claim is supported by the comparison of M\,4 and NGC\,2808 by \citet[][section 4]{Milone2014}.
It is worthwhile to consider the value of this technique in light of the current capabilities of \emph{HST} WFC3/IR as well as the 
advent of the \emph{James Webb Space Telescope}.

\subsection{Helium Abundance Variation from Multiband Photometry of the Main Sequence}
\citet{Milone2013} used the broad wavelength coverage of the \emph{HST} photometry, along with the ATLAS and SYNTHE
codes, to generate model atmospheres and synthetic spectra for the stellar populations in NGC\,6752. They compared
the observed colour difference between populations A and C at fixed $F814W$ magnitude in several colours with synthetic spectra. 
The synthetic spectra for population C included varying helium abundance (up to Y=0.29) and \citet{Milone2013} used these 
to estimate the helium abundance of population C: an enhancement above population A by $\dY \sim 0.03$.

The same analysis can be performed with the isochrones at the location on the MS adopted by Milone et al., $F814W=18.5$.
To do so, the magnitudes of 8 filters considered by \citet[][see their Figure 9]{Milone2013} were extracted from the isochrones
at $F814W=18.5$. The colour difference between cases A and C, given by $(Filter - F814W)_\mathrm{A} -(Filter - F814W)_\mathrm{C}$, was 
computed for each filter considered; the same was done for the difference between cases A and $\CdY$. The results of this
exercise are plotted in Figure \ref{fig:MSY}, where the two variants of case C are identified by their level of helium-enhancement 
relative to case A; this is done to highlight the influence of helium abundance on the MS colours at fixed $F814W$ magnitude.

The UV filters have been excluded from Figure \ref{fig:MSY} because they are the most
sensitive to variations in light elements other than helium. However, starting with the $F390M$ filter and extending
redward, the ATLAS/SYNTHE isochrone for case $\CdY$ ($\dY=0.03$) provide a good match to the observations. This should
not come as a surprise because the level of helium-enhancement estimated by \citet{Milone2013} was based on comparisons
of their data with ATLAS/SYNTHE models. Nor can it be taken as an independent confirmation of that result because our case $\CdY$
was chosen based on the estimate from \citet{Milone2013}. Having said all that, the left-hand panel of Figure \ref{fig:MSY} is a 
confirmation that some degree of helium-enhancement is required to match the observation because the case C isochrone ($\dY=0$) shows
considerably less variation relative to case A, which is inconsistent with the data.

The right-hand panel of Figure \ref{fig:MSY} provides an important check on this technique. Whereas the ATLAS/SYNTHE 
isochrones produced in this study are consistent with the estimate of helium-enhancement using atmosphere models from (essentially) 
the same codes, the \phx\ models would require a larger helium-enhancement (by a factor of $\la2$) in population C in order to match 
the observations. We conclude that the use of multiband photometry to estimate helium variation among different populations in GCs is
a viable technique but that the resulting $\dY$ depends on the model atmospheres employed. More work in this direction
is certainly called for.

\section{Conclusions}\label{conclusions}
In the context of multiple population in GCs, this study represents the first successful 
confrontation between the best available observational material, both photometry and spectroscopy, and 
self-consistent theoretical modelling of stellar interiors and atmospheres.

This study presents self-consistent stellar atmosphere and evolution models directly addressing the presence
of multiple stellar populations on the MS of NGC\,6752. The abundances adopted in the models are
taken from spectroscopic analyses. Two sets of model atmospheres and synthetic spectra were computed using
ATLAS/SYNTHE and \phx\ in order to reveal the differences that may arise from adopting one set of atmosphere
models or another. 

\comment{
ATLAS/SYNTHE and \phx\ spectra differ somewhat, particularly in the UV, 
which translates into noticeable differences in the UV and blue synthetic CMDs.
It is not obvious that one set of models matches the \emph{HST} $F275W$ data better 
than the other, though the ATLAS/SYNTHE models perform better in $F336W-F410M$.} 
Further improvements to the model atmospheres and synthetic spectra 
that directly address the UV would be welcome. 

The combination of a good match between the models and the optical CMDs and the fact that multiple 
populations are essentially indistinguishable in the optical CMDs recommends the use of optical CMDs for GC 
age analyses. This point does not extend to those GCs whose optical CMDs reveal the presence of 
multiple populations \citep[e.g., NGC 1851;][]{Milone2008}.

The spread of the lower MS in the near-IR CMD, caused by water absorption, provides a useful probe of
the range of oxygen-variation in the cluster provided that photometric errors are accounted for and stellar
models for the appropriate compositions are used. \comment{\phx\ and ATLAS/SYNTHE models predict essentially the 
same behavior within $\sim1$ magnitude of the `knee' though the ATLAS/SYNTHE models trace the knee itself
better.} This feature represents a relatively inexpensive
way to estimate the level of oxygen-variation in a GC.

The models support the use of multiband photometry to estimate helium variations among the different populations
in NGC\,6752 and other, similar GCs using the technique developed by Milone and collaborators. However, it
is important to note that the estimate of helium variation is model-dependent: ATLAS/SYNTHE models show a
stronger sensitivity to helium variation than \phx\ models.

The next step in this project will be to extend the current analysis to the RGB and the HB.
The \citet{Milone2013} photometry covers the full extent of the RGB and HB stars and, through detailed modeling,
we intend to study the complex relationship between light-element abundance variations (including helium) and 
cumulative mass loss on the RGB as manifested in the HB morphology.

The phenomenon of multiple stellar populations in GCs represents a major challenge to our understanding 
of stellar evolution and nucleosynthesis. Successfully reproducing the observed characteristics at all 
evolutionary stages, from the MS through the RGB and onto the HB, would serve as an important validation 
of stellar physics and its wider application.

\section*{Acknowledgments}
We thank R.\ Kurucz for allowing us to use the latest version of his codes and line lists in advance of publication.
AD received support from the Australian Research Council under grant FL110100012.
CC received support from NSF grant AST-1313280 and NASA grant NNX13AI46G.


\begin{thebibliography}{52}
\expandafter\ifx\csname natexlab\endcsname\relax\def\natexlab#1{#1}\fi

\bibitem[{{Anderson}(1997)}]{JayPhD}
{Anderson}, A.~J. 1997, PhD thesis, UNIVERSITY OF CALIFORNIA, BERKELEY

\bibitem[{{Anderson} {et~al.}(2008){Anderson}, {Sarajedini}, {Bedin}, {King},
  {Piotto}, {Reid}, {Siegel}, {Majewski}, {Paust}, {Aparicio}, {Milone},
  {Chaboyer}, \& {Rosenberg}}]{Anderson2008}
{Anderson}, J., {Sarajedini}, A., {Bedin}, L.~R., {King}, I.~R., {Piotto}, G.,
  {Reid}, I.~N., {Siegel}, M., {Majewski}, S.~R., {Paust}, N.~E.~Q.,
  {Aparicio}, A., {Milone}, A.~P., {Chaboyer}, B., \& {Rosenberg}, A. 2008, AJ,
  135, 2055

\bibitem[{{Asplund} {et~al.}(2009){Asplund}, {Grevesse}, {Sauval}, \&
  {Scott}}]{Asplund2009}
{Asplund}, M., {Grevesse}, N., {Sauval}, A.~J., \& {Scott}, P. 2009, ARA\&A,
  47, 481

\bibitem[{{Bedin} {et~al.}(2004){Bedin}, {Piotto}, {Anderson}, {Cassisi},
  {King}, {Momany}, \& {Carraro}}]{Bedin2004}
{Bedin}, L.~R., {Piotto}, G., {Anderson}, J., {Cassisi}, S., {King}, I.~R.,
  {Momany}, Y., \& {Carraro}, G. 2004, ApJL, 605, L125

\bibitem[{{Cardelli} {et~al.}(1989){Cardelli}, {Clayton}, \& {Mathis}}]{CCM}
{Cardelli}, J.~A., {Clayton}, G.~C., \& {Mathis}, J.~S. 1989, ApJ, 345, 245

\bibitem[{{Carretta} {et~al.}(2009{\natexlab{a}}){Carretta}, {Bragaglia},
  {Gratton}, \& {Lucatello}}]{Carretta2009A}
{Carretta}, E., {Bragaglia}, A., {Gratton}, R., \& {Lucatello}, S.
  2009{\natexlab{a}}, A\&A, 505, 139

\bibitem[{{Carretta} {et~al.}(2009{\natexlab{b}}){Carretta}, {Bragaglia},
  {Gratton}, {Lucatello}, {Catanzaro}, {Leone}, {Bellazzini}, {Claudi},
  {D'Orazi}, {Momany}, {Ortolani}, {Pancino}, {Piotto}, {Recio-Blanco}, \&
  {Sabbi}}]{Carretta2009B}
{Carretta}, E., {Bragaglia}, A., {Gratton}, R.~G., {Lucatello}, S.,
  {Catanzaro}, G., {Leone}, F., {Bellazzini}, M., {Claudi}, R., {D'Orazi}, V.,
  {Momany}, Y., {Ortolani}, S., {Pancino}, E., {Piotto}, G., {Recio-Blanco},
  A., \& {Sabbi}, E. 2009{\natexlab{b}}, A\&A, 505, 117

\bibitem[{{Carretta} {et~al.}(2005){Carretta}, {Gratton}, {Lucatello},
  {Bragaglia}, \& {Bonifacio}}]{Carretta2005}
{Carretta}, E., {Gratton}, R.~G., {Lucatello}, S., {Bragaglia}, A., \&
  {Bonifacio}, P. 2005, A\&A, 433, 597

\bibitem[{{di Criscienzo} {et~al.}(2010){di Criscienzo}, {D'Antona}, {Ventura},
  \& {}}]{diCriscienzo2010A}
{di Criscienzo}, M., {D'Antona}, F., {Ventura}, P., \& {}. 2010, A\&A, 511, A70

\bibitem[{{Dotter} {et~al.}(2007{\natexlab{a}}){Dotter}, {Chaboyer},
  {Ferguson}, {Lee}, {Worthey}, {Jevremovi{\'c}}, \& {Baron}}]{Dotter2007}
{Dotter}, A., {Chaboyer}, B., {Ferguson}, J.~W., {Lee}, H.-c., {Worthey}, G.,
  {Jevremovi{\'c}}, D., \& {Baron}, E. 2007{\natexlab{a}}, ApJ, 666, 403

\bibitem[{{Dotter} {et~al.}(2007{\natexlab{b}}){Dotter}, {Chaboyer},
  {Jevremovi{\'c}}, {Baron}, {Ferguson}, {Sarajedini}, \&
  {Anderson}}]{Dotter2007B}
{Dotter}, A., {Chaboyer}, B., {Jevremovi{\'c}}, D., {Baron}, E., {Ferguson},
  J.~W., {Sarajedini}, A., \& {Anderson}, J. 2007{\natexlab{b}}, AJ, 134, 376

\bibitem[{{Dotter} {et~al.}(2008){Dotter}, {Chaboyer}, {Jevremovi{\'c}},
  {Kostov}, {Baron}, \& {Ferguson}}]{Dotter2008}
{Dotter}, A., {Chaboyer}, B., {Jevremovi{\'c}}, D., {Kostov}, V., {Baron}, E.,
  \& {Ferguson}, J.~W. 2008, ApJS, 178, 89

\bibitem[{{Ferguson} {et~al.}(2005){Ferguson}, {Alexander}, {Allard}, {Barman},
  {Bodnarik}, {Hauschildt}, {Heffner-Wong}, \& {Tamanai}}]{Ferguson2005}
{Ferguson}, J.~W., {Alexander}, D.~R., {Allard}, F., {Barman}, T., {Bodnarik},
  J.~G., {Hauschildt}, P.~H., {Heffner-Wong}, A., \& {Tamanai}, A. 2005, ApJ,
  623, 585

\bibitem[{{Freeman} \& {Norris}(1981)}]{FreemanNorris1981}
{Freeman}, K.~C. \& {Norris}, J. 1981, ARA\&A, 19, 319

\bibitem[{{Girardi} {et~al.}(2007){Girardi}, {Castelli}, {Bertelli}, \&
  {Nasi}}]{Girardi2007}
{Girardi}, L., {Castelli}, F., {Bertelli}, G., \& {Nasi}, E. 2007, A\&A, 468,
  657

\bibitem[{{Gratton} {et~al.}(2004){Gratton}, {Sneden}, {Carretta}, \&
  {}}]{Gratton2004}
{Gratton}, R., {Sneden}, C., {Carretta}, E., \& {}. 2004, ARA\&A, 42, 385

\bibitem[{{Gratton} {et~al.}(2012){Gratton}, {Carretta}, \&
  {Bragaglia}}]{Gratton2012}
{Gratton}, R.~G., {Carretta}, E., \& {Bragaglia}, A. 2012, A\&ARv, 20, 50

\bibitem[{{Grundahl} {et~al.}(2002){Grundahl}, {Briley}, {Nissen}, \&
  {Feltzing}}]{Grundahl2002}
{Grundahl}, F., {Briley}, M., {Nissen}, P.~E., \& {Feltzing}, S. 2002, A\&Aa,
  385, L14

\bibitem[{{Harris}(1996)}]{Harris1996}
{Harris}, W.~E. 1996, AJ, 112, 1487

\bibitem[{{Hauschildt} {et~al.}(1999{\natexlab{a}}){Hauschildt}, {Allard},
  {Baron}, \& {}e}]{Hauschildt1999A}
{Hauschildt}, P.~H., {Allard}, F., {Baron}, E., \& {}e. 1999{\natexlab{a}},
  ApJ, 512, 377

\bibitem[{{Hauschildt} {et~al.}(1999{\natexlab{b}}){Hauschildt}, {Allard},
  {Ferguson}, {Baron}, \& {Alexander}}]{Hauschildt1999B}
{Hauschildt}, P.~H., {Allard}, F., {Ferguson}, J., {Baron}, E., \& {Alexander},
  D.~R. 1999{\natexlab{b}}, ApJ, 525, 871

\bibitem[{{Iglesias} \& {Rogers}(1996)}]{OPAL1996}
{Iglesias}, C.~A. \& {Rogers}, F.~J. 1996, ApJ, 464, 943

\bibitem[{{Kalirai} {et~al.}(2012){Kalirai}, {Richer}, {Anderson}, {Dotter},
  {Fahlman}, {Hansen}, {Hurley}, {King}, {Reitzel}, {Rich}, {Shara}, {Stetson},
  \& {Woodley}}]{Kalirai2012}
{Kalirai}, J.~S., {Richer}, H.~B., {Anderson}, J., {Dotter}, A., {Fahlman},
  G.~G., {Hansen}, B.~M.~S., {Hurley}, J., {King}, I.~R., {Reitzel}, D.,
  {Rich}, R.~M., {Shara}, M.~M., {Stetson}, P.~B., \& {Woodley}, K.~A. 2012,
  AJ, 143, 11

\bibitem[{{Kurucz}(1970)}]{Kurucz1970}
{Kurucz}, R.~L. 1970, SAO Special Report, 309

\bibitem[{{Kurucz}(1993)}]{Kurucz1993}
---. 1993, {SYNTHE spectrum synthesis programs and line data} (CD-ROM)

\bibitem[{{Kurucz} \& {Avrett}(1981)}]{Kurucz1981}
{Kurucz}, R.~L. \& {Avrett}, E.~H. 1981, SAO Special Report, 391

\bibitem[{{Lee} {et~al.}(1999){Lee}, {Joo}, {Sohn}, {Rey}, {Lee}, \&
  {Walker}}]{Lee1999}
{Lee}, Y.-W., {Joo}, J.-M., {Sohn}, Y.-J., {Rey}, S.-C., {Lee}, H.-C., \&
  {Walker}, A.~R. 1999, Nature, 402, 55

\bibitem[{{Marino} {et~al.}(2011){Marino}, {Villanova}, {Milone}, {Piotto},
  {Lind}, {Geisler}, \& {Stetson}}]{Marino2011}
{Marino}, A.~F., {Villanova}, S., {Milone}, A.~P., {Piotto}, G., {Lind}, K.,
  {Geisler}, D., \& {Stetson}, P.~B. 2011, ApJL, 730, L16

\bibitem[{{Marino} {et~al.}(2008){Marino}, {Villanova}, {Piotto}, {Milone},
  {Momany}, {Bedin}, \& {Medling}}]{Marino2008}
{Marino}, A.~F., {Villanova}, S., {Piotto}, G., {Milone}, A.~P., {Momany}, Y.,
  {Bedin}, L.~R., \& {Medling}, A.~M. 2008, A\&A, 490, 625

\bibitem[{{Milone} {et~al.}(2008){Milone}, {Bedin}, {Piotto}, {Anderson},
  {King}, {Sarajedini}, {Dotter}, {Chaboyer}, {Mar{\'{\i}}n-Franch},
  {Majewski}, {Aparicio}, {Hempel}, {Paust}, {Reid}, {Rosenberg}, \&
  {Siegel}}]{Milone2008}
{Milone}, A.~P., {Bedin}, L.~R., {Piotto}, G., {Anderson}, J., {King}, I.~R.,
  {Sarajedini}, A., {Dotter}, A., {Chaboyer}, B., {Mar{\'{\i}}n-Franch}, A.,
  {Majewski}, S., {Aparicio}, A., {Hempel}, M., {Paust}, N.~E.~Q., {Reid},
  I.~N., {Rosenberg}, A., \& {Siegel}, M. 2008, ApJ, 673, 241

\bibitem[{{Milone} {et~al.}(2014){Milone}, {Marino}, {Bedin}, {Piotto},
  {Cassisi}, {Dieball}, {Anderson}, {Jerjen}, {Asplund}, {Bellini}, {Brogaard},
  {Dotter}, {Giersz}, {Heggie}, {Knigge}, {Rich}, {van den Berg}, \&
  {Buonanno}}]{Milone2014}
{Milone}, A.~P., {Marino}, A.~F., {Bedin}, L.~R., {Piotto}, G., {Cassisi}, S.,
  {Dieball}, A., {Anderson}, J., {Jerjen}, H., {Asplund}, M., {Bellini}, A.,
  {Brogaard}, K., {Dotter}, A., {Giersz}, M., {Heggie}, D.~C., {Knigge}, C.,
  {Rich}, R.~M., {van den Berg}, M., \& {Buonanno}, R. 2014, MNRAS, 439, 1588

\bibitem[{{Milone} {et~al.}(2012){Milone}, {Marino}, {Cassisi}, {Piotto},
  {Bedin}, {Anderson}, {Allard}, {Aparicio}, {Bellini}, {Buonanno}, {Monelli},
  \& {Pietrinferni}}]{Milone2012}
{Milone}, A.~P., {Marino}, A.~F., {Cassisi}, S., {Piotto}, G., {Bedin}, L.~R.,
  {Anderson}, J., {Allard}, F., {Aparicio}, A., {Bellini}, A., {Buonanno}, R.,
  {Monelli}, M., \& {Pietrinferni}, A. 2012, ApJL, 754, L34

\bibitem[{{Milone} {et~al.}(2013){Milone}, {Marino}, {Piotto}, {Bedin},
  {Anderson}, {Aparicio}, {Bellini}, {Cassisi}, {D'Antona}, {Grundahl},
  {Monelli}, \& {Yong}}]{Milone2013}
{Milone}, A.~P., {Marino}, A.~F., {Piotto}, G., {Bedin}, L.~R., {Anderson}, J.,
  {Aparicio}, A., {Bellini}, A., {Cassisi}, S., {D'Antona}, F., {Grundahl}, F.,
  {Monelli}, M., \& {Yong}, D. 2013, ApJ, 767, 120

\bibitem[{{Milone} {et~al.}(2010){Milone}, {Piotto}, {King}, {Bedin},
  {Anderson}, {Marino}, {Momany}, {Malavolta}, \& {Villanova}}]{Milone2010}
{Milone}, A.~P., {Piotto}, G., {King}, I.~R., {Bedin}, L.~R., {Anderson}, J.,
  {Marino}, A.~F., {Momany}, Y., {Malavolta}, L., \& {Villanova}, S. 2010, ApJ,
  709, 1183

\bibitem[{{Pietrinferni} {et~al.}(2009){Pietrinferni}, {Cassisi}, {Salaris},
  {Percival}, \& {Ferguson}}]{Pietrinferni2009}
{Pietrinferni}, A., {Cassisi}, S., {Salaris}, M., {Percival}, S., \&
  {Ferguson}, J.~W. 2009, ApJ, 697, 275

\bibitem[{{Piotto}(2009)}]{Piotto2009}
{Piotto}, G. 2009, in IAU Symposium, Vol. 258, IAU Symposium, ed. E.~E.
  {Mamajek}, D.~R. {Soderblom}, \& R.~F.~G. {Wyse}, 233--244

\bibitem[{{Prantzos} {et~al.}(2007){Prantzos}, {Charbonnel}, {Iliadis}, \&
  {}}]{Prantzos2007}
{Prantzos}, N., {Charbonnel}, C., {Iliadis}, C., \& {}. 2007, A\&A, 470, 179

\bibitem[{{Richer} {et~al.}(2008){Richer}, {Dotter}, {Hurley}, {Anderson},
  {King}, {Davis}, {Fahlman}, {Hansen}, {Kalirai}, {Paust}, {Rich}, \&
  {Shara}}]{Richer2008}
{Richer}, H.~B., {Dotter}, A., {Hurley}, J., {Anderson}, J., {King}, I.,
  {Davis}, S., {Fahlman}, G.~G., {Hansen}, B.~M.~S., {Kalirai}, J., {Paust},
  N., {Rich}, R.~M., \& {Shara}, M.~M. 2008, AJ, 135, 2141

\bibitem[{{Salaris} {et~al.}(2006){Salaris}, {Weiss}, {Ferguson}, \&
  {Fusilier}}]{Salaris2006}
{Salaris}, M., {Weiss}, A., {Ferguson}, J.~W., \& {Fusilier}, D.~J. 2006, ApJ,
  645, 1131

\bibitem[{{Sarajedini} {et~al.}(2007){Sarajedini}, {Bedin}, {Chaboyer},
  {Dotter}, {Siegel}, {Anderson}, {Aparicio}, {King}, {Majewski},
  {Mar{\'{\i}}n-Franch}, {Piotto}, {Reid}, \& {Rosenberg}}]{Sarajedini2007}
{Sarajedini}, A., {Bedin}, L.~R., {Chaboyer}, B., {Dotter}, A., {Siegel}, M.,
  {Anderson}, J., {Aparicio}, A., {King}, I., {Majewski}, S.,
  {Mar{\'{\i}}n-Franch}, A., {Piotto}, G., {Reid}, I.~N., \& {Rosenberg}, A.
  2007, AJ, 133, 1658

\bibitem[{{Sbordone} {et~al.}(2004){Sbordone}, {Bonifacio}, {Castelli}, \&
  {Kurucz}}]{Sbordone2004}
{Sbordone}, L., {Bonifacio}, P., {Castelli}, F., \& {Kurucz}, R.~L. 2004,
  Memorie della Societa Astronomica Italiana Supplementi, 5, 93

\bibitem[{{Sbordone} {et~al.}(2011){Sbordone}, {Salaris}, {Weiss}, \&
  {Cassisi}}]{Sbordone2011}
{Sbordone}, L., {Salaris}, M., {Weiss}, A., \& {Cassisi}, S. 2011, A\&A, 534,
  A9

\bibitem[{{Schlafly} \& {Finkbeiner}(2011)}]{SF11}
{Schlafly}, E.~F. \& {Finkbeiner}, D.~P. 2011, ApJ, 737, 103

\bibitem[{{Smith} \& {Briley}(2005)}]{Smith2005B}
{Smith}, G.~H. \& {Briley}, M.~M. 2005, PASP, 117, 895

\bibitem[{{Smith} \& {Briley}(2006)}]{Smith2006}
---. 2006, PASP, 118, 740

\bibitem[{{Smith} {et~al.}(2005){Smith}, {Briley}, {Harbeck}, \&
  {}}]{Smith2005A}
{Smith}, G.~H., {Briley}, M.~M., {Harbeck}, D., \& {}. 2005, AJ, 129, 1589

\bibitem[{{VandenBerg} {et~al.}(2012){VandenBerg}, {Bergbusch}, {Dotter},
  {Ferguson}, {Michaud}, {Richer}, \& {Proffitt}}]{VandenBerg2012}
{VandenBerg}, D.~A., {Bergbusch}, P.~A., {Dotter}, A., {Ferguson}, J.~W.,
  {Michaud}, G., {Richer}, J., \& {Proffitt}, C.~R. 2012, ApJ, 755, 15

\bibitem[{{Villanova} {et~al.}(2009){Villanova}, {Piotto}, {Gratton}, \&
  {}}]{Villanova2009}
{Villanova}, S., {Piotto}, G., {Gratton}, R.~G., \& {}. 2009, A\&A, 499, 755

\bibitem[{{Yong} {et~al.}(2008){Yong}, {Grundahl}, {Johnson}, \&
  {Asplund}}]{Yong2008}
{Yong}, D., {Grundahl}, F., {Johnson}, J.~A., \& {Asplund}, M. 2008, ApJ, 684,
  1159

\bibitem[{{Yong} {et~al.}(2003){Yong}, {Grundahl}, {Lambert}, {Nissen}, \&
  {Shetrone}}]{Yong2003}
{Yong}, D., {Grundahl}, F., {Lambert}, D.~L., {Nissen}, P.~E., \& {Shetrone},
  M.~D. 2003, A\&A, 402, 985

\bibitem[{{Yong} {et~al.}(2005){Yong}, {Grundahl}, {Nissen}, {Jensen}, \&
  {Lambert}}]{Yong2005}
{Yong}, D., {Grundahl}, F., {Nissen}, P.~E., {Jensen}, H.~R., \& {Lambert},
  D.~L. 2005, A\&A, 438, 875

\bibitem[{{Yong} {et~al.}(2013){Yong}, {Mel{\'e}ndez}, {Grundahl}, {Roederer},
  {Norris}, {Milone}, {Marino}, {Coelho}, {McArthur}, {Lind}, {Collet}, \&
  {Asplund}}]{Yong2013}
{Yong}, D., {Mel{\'e}ndez}, J., {Grundahl}, F., {Roederer}, I.~U., {Norris},
  J.~E., {Milone}, A.~P., {Marino}, A.~F., {Coelho}, P., {McArthur}, B.~E.,
  {Lind}, K., {Collet}, R., \& {Asplund}, M. 2013, MNRAS, 434, 3542

\end{thebibliography}

\appendix

\section{Abundances for Model Atmospheres and Opacities}\label{app}

The abundances of the elements assumed in the models, determined from the abundance ratios listed in 
\ref{tab:ratio}, for cases A and C are recorded here. The reference solar scale is that of \citet{Asplund2009}.
The $\log(\mathrm{A})$, number fractions, and mass fractions input to the stellar atmosphere codes, including 
low-temperature opacities, are given in Table \ref{tab:abund}. The number fractions for the
23 elements input to the OPAL web server are given in Table \ref{tab:opal}.

\begin{table*}
  \centering
  \caption{Abundances used in model atmospheres, low-temperature opacities, and synthetic spectra\label{tab:abund}}
  \begin{tabular}{lrrrrrr}
    \hline
    &  \multicolumn{3}{c}{Population A} & \multicolumn{3}{c}{Population C} \\
 El & $\log$(A) &Num.\ Frac.&Mass Frac.&$\log$(A)&Num.\ Frac.&Mass Frac.\\
 \hline
  H  & $ 12.00$  &9.215E-01&  7.467E-01  &$  12.00$  &9.215E-01 & 7.468E-01\\
  He & $ 10.93$  &7.843E-02&  2.523E-01  &$  10.93$  &7.843E-02 & 2.524E-01\\
  Li & $  1.61$  &3.754E-11&  2.095E-10  &$   1.65$  &4.116E-11 & 2.297E-10\\ 
  Be & $ -0.27$  &4.948E-13&  3.585E-12  &$  -0.23$  &5.426E-13 & 3.931E-12\\ 
  B  & $  1.05$  &1.033E-11&  8.987E-11  &$   1.09$  &1.133E-11 & 9.855E-11\\
  C  & $  6.53$  &3.122E-06&  3.015E-05  &$   6.12$  &1.214E-06 & 1.173E-05\\
  N  & $  6.07$  &1.082E-06&  1.219E-05  &$   7.57$  &3.423E-05 & 3.855E-04\\ 
  O  & $  7.69$  &4.513E-05&  5.805E-04  &$   7.11$  &1.187E-05 & 1.527E-04\\
  F  & $  2.91$  &7.490E-10&  1.144E-08  &$   2.95$  &8.212E-10 & 1.254E-08\\
  Ne & $  6.68$  &4.410E-06&  7.152E-05  &$   6.72$  &4.836E-06 & 7.843E-05\\
  Na & $  4.56$  &3.345E-08&  6.184E-07  &$   5.24$  &1.601E-07 & 2.960E-06\\
  Mg & $  6.46$  &2.657E-06&  5.194E-05  &$   6.39$  &2.262E-06 & 4.421E-05\\
  Al & $  5.08$  &1.107E-07&  2.403E-06  &$   5.98$  &8.800E-07 & 1.909E-05\\
  Si & $  6.13$  &1.243E-06&  2.806E-05  &$   6.25$  &1.638E-06 & 3.700E-05\\
  P  & $  3.76$  &5.302E-09&  1.320E-07  &$   3.80$  &5.814E-09 & 1.447E-07\\
  S  & $  5.72$  &4.836E-07&  1.246E-05  &$   5.76$  &5.302E-07 & 1.367E-05\\
  Cl & $  3.85$  &6.523E-09&  1.859E-07  &$   3.89$  &7.153E-09 & 2.039E-07\\
  Ar & $  4.75$  &5.182E-08&  1.664E-06  &$   4.79$  &5.681E-08 & 1.824E-06\\
  K  & $  3.38$  &2.210E-09&  6.949E-08  &$   3.42$  &2.423E-09 & 7.619E-08\\
  Ca & $  4.90$  &7.319E-08&  2.358E-06  &$   5.00$  &9.215E-08 & 2.969E-06\\
  Sc & $  1.50$  &2.914E-11&  1.053E-09  &$   1.54$  &3.195E-11 & 1.154E-09\\
  Ti & $  3.40$  &2.314E-09&  8.910E-08  &$   3.49$  &2.847E-09 & 1.096E-07\\
  V  & $  1.94$  &8.025E-11&  3.287E-09  &$   2.07$  &1.082E-10 & 4.434E-09\\
  Cr & $  3.99$  &9.005E-09&  3.764E-07  &$   4.03$  &9.874E-09 & 4.128E-07\\
  Mn & $  3.28$  &1.755E-09&  7.755E-08  &$   3.37$  &2.160E-09 & 9.541E-08\\
  Fe & $  5.85$  &6.523E-07&  2.929E-05  &$   5.89$  &7.153E-07 & 3.211E-05\\
  Co & $  3.31$  &1.881E-09&  8.914E-08  &$   3.32$  &1.925E-09 & 9.122E-08\\
  Ni & $  4.51$  &2.981E-08&  1.407E-06  &$   4.58$  &3.503E-08 & 1.653E-06\\
  Cu & $  1.88$  &6.990E-11&  3.571E-09  &$   1.98$  &8.800E-11 & 4.496E-09\\
  Zn & $  2.91$  &7.490E-10&  3.937E-08  &$   2.95$  &8.212E-10 & 4.317E-08\\
  Ga & $  1.39$  &2.262E-11&  1.267E-09  &$   1.43$  &2.480E-11 & 1.390E-09\\
  Ge & $  2.00$  &9.215E-11&  5.380E-09  &$   2.04$  &1.010E-10 & 5.900E-09\\
  As & $  0.65$  &4.116E-12&  2.479E-10  &$   0.69$  &4.513E-12 & 2.718E-10\\
  Se & $  1.69$  &4.513E-11&  2.866E-09  &$   1.73$  &4.948E-11 & 3.142E-09\\
  Br & $  0.89$  &7.153E-12&  4.595E-10  &$   0.93$  &7.843E-12 & 5.038E-10\\
  Kr & $  1.60$  &3.668E-11&  2.471E-09  &$   1.64$  &4.022E-11 & 2.710E-09\\
  Rb & $  0.87$  &6.831E-12&  4.694E-10  &$   0.91$  &7.490E-12 & 5.147E-10\\
  Sr & $  1.22$  &1.529E-11&  1.077E-09  &$   1.26$  &1.676E-11 & 1.181E-09\\
  Y  & $  0.56$  &3.345E-12&  2.391E-10  &$   0.60$  &3.668E-12 & 2.622E-10\\
  Zr & $  0.93$  &7.843E-12&  5.752E-10  &$   0.97$  &8.600E-12 & 6.307E-10\\
  Nb & $ -0.19$  &5.949E-13&  4.444E-11  &$  -0.15$  &6.523E-13 & 4.873E-11\\
  Mo & $  0.23$  &1.564E-12&  1.206E-10  &$   0.27$  &1.715E-12 & 1.322E-10\\
  Ru & $  0.10$  &1.160E-12&  9.426E-11  &$   0.14$  &1.272E-12 & 1.033E-10\\
  Rh & $ -0.74$  &1.676E-13&  1.387E-11  &$  -0.70$  &1.838E-13 & 1.521E-11\\
  Pd & $ -0.08$  &7.664E-13&  6.558E-11  &$  -0.04$  &8.404E-13 & 7.191E-11\\
  Ag & $ -0.71$  &1.796E-13&  1.558E-11  &$  -0.67$  &1.970E-13 & 1.708E-11\\
  Cd & $  0.06$  &1.058E-12&  9.563E-11  &$   0.10$  &1.160E-12 & 1.048E-10\\
  In & $ -0.85$  &1.301E-13&  1.201E-11  &$  -0.81$  &1.427E-13 & 1.317E-11\\
  Sn & $  0.39$  &2.262E-12&  2.159E-10  &$   0.43$  &2.480E-12 & 2.367E-10\\
  Sb & $ -0.64$  &2.111E-13&  2.066E-11  &$  -0.60$  &2.314E-13 & 2.266E-11\\
  Te & $  0.53$  &3.122E-12&  3.203E-10  &$   0.57$  &3.423E-12 & 3.513E-10\\
  I  & $ -0.10$  &7.319E-13&  7.468E-11  &$  -0.06$  &8.026E-13 & 8.189E-11\\
  Xe & $  0.59$  &3.585E-12&  3.784E-10  &$   0.63$  &3.930E-12 & 4.149E-10\\
  \hline
   \end{tabular}
 \end{table*}

\begin{table*}
  \centering
  \contcaption{}
  \begin{tabular}{lrrrrrr}
    \hline
    &\multicolumn{3}{c}{Population A} & \multicolumn{3}{c}{Population C} \\
    &         & \multicolumn{2}{c}{Fraction} & & \multicolumn{2}{c}{Fraction} \\
 El &$\log$(A) &By\ Number &  By\ Mass &$\log$(A)&By Number &  By Mass     \\
 \hline
  Cs & $ -0.57$  &2.480E-13&  2.650E-11  &$  -0.53$  &2.719E-13 & 2.906E-11\\
  Ba & $  0.53$  &3.122E-12&  3.447E-10  &$   0.57$  &3.423E-12 & 3.780E-10\\
  La & $ -0.55$  &2.597E-13&  2.900E-11  &$  -0.51$  &2.847E-13 & 3.180E-11\\
  Ce & $ -0.07$  &7.843E-13&  8.835E-11  &$  -0.03$  &8.600E-13 & 9.688E-11\\
  Pr & $ -0.93$  &1.082E-13&  1.226E-11  &$  -0.89$  &1.187E-13 & 1.344E-11\\
  Nd & $ -0.23$  &5.426E-13&  6.292E-11  &$  -0.19$  &5.949E-13 & 6.899E-11\\
  Sm & $ -0.69$  &1.881E-13&  2.274E-11  &$  -0.65$  &2.063E-13 & 2.494E-11\\
  Eu & $ -1.13$  &6.831E-14&  8.346E-12  &$  -1.09$  &7.490E-14 & 9.151E-12\\
  Gd & $ -0.58$  &2.423E-13&  3.064E-11  &$  -0.54$  &2.657E-13 & 3.360E-11\\
  Tb & $ -1.35$  &4.116E-14&  5.259E-12  &$  -1.31$  &4.513E-14 & 5.767E-12\\
  Dy & $ -0.55$  &2.597E-13&  3.393E-11  &$  -0.51$  &2.847E-13 & 3.720E-11\\
  Ho & $ -1.17$  &6.230E-14&  8.261E-12  &$  -1.13$  &6.831E-14 & 9.058E-12\\
  Er & $ -0.73$  &1.715E-13&  2.307E-11  &$  -0.69$  &1.881E-13 & 2.530E-11\\
  Tm & $ -1.55$  &2.597E-14&  3.527E-12  &$  -1.51$  &2.847E-14 & 3.867E-12\\
  Yb & $ -0.81$  &1.427E-13&  1.985E-11  &$  -0.77$  &1.564E-13 & 2.177E-11\\
  Lu & $ -1.55$  &2.597E-14&  3.653E-12  &$  -1.51$  &2.847E-14 & 4.006E-12\\
  Hf & $ -0.80$  &1.460E-13&  2.095E-11  &$  -0.76$  &1.601E-13 & 2.298E-11\\
  Ta & $ -1.77$  &1.564E-14&  2.276E-12  &$  -1.73$  &1.715E-14 & 2.496E-12\\
  W  & $ -0.80$  &1.460E-13&  2.158E-11  &$  -0.76$  &1.601E-13 & 2.367E-11\\
  Re & $ -1.39$  &3.754E-14&  5.620E-12  &$  -1.35$  &4.116E-14 & 6.162E-12\\
  Os & $ -0.25$  &5.182E-13&  7.925E-11  &$  -0.21$  &5.681E-13 & 8.690E-11\\
  Ir & $ -0.27$  &4.948E-13&  7.647E-11  &$  -0.23$  &5.426E-13 & 8.385E-11\\
  Pt & $ -0.03$  &8.600E-13&  1.348E-10  &$   0.01$  &9.429E-13 & 1.479E-10\\
  Au & $ -0.73$  &1.715E-13&  2.717E-11  &$  -0.69$  &1.881E-13 & 2.979E-11\\
  Hg & $ -0.48$  &3.051E-13&  4.921E-11  &$  -0.44$  &3.345E-13 & 5.396E-11\\
  Tl & $ -0.75$  &1.638E-13&  2.692E-11  &$  -0.71$  &1.796E-13 & 2.952E-11\\
  Pb & $  0.10$  &1.160E-12&  1.932E-10  &$   0.14$  &1.272E-12 & 2.119E-10\\
  Bi & $ -1.00$  &9.215E-14&  1.548E-11  &$  -0.96$  &1.010E-13 & 1.697E-11\\
  Th & $ -1.63$  &2.160E-14&  4.029E-12  &$  -1.59$  &2.368E-14 & 4.418E-12\\
  U  & $ -2.19$  &5.949E-15&  1.138E-12  &$  -2.15$  &6.523E-15 & 1.248E-12\\
  \hline                                          
   \end{tabular}                                  
 \end{table*}

\begin{table*}
 \centering
 \caption{Abundances used in the OPAL opacities\label{tab:opal}}
 \begin{tabular}{lrr}
 \hline
    & Population A&Population C \\
 EL & Num.\ Frac. & Num.\ Frac. \\
 \hline
  C  &  0.052819 &  0.020743\\ 
  N  &  0.018314 &  0.584609\\
  O  &  0.763472 &  0.202705\\
  F  &  0.000013 &  0.000014\\
  Ne &  0.074609 &  0.082578\\
  Na &  0.000566 &  0.002734\\
  Mg &  0.044957 &  0.038625\\
  Al &  0.001874 &  0.015027\\
  Si &  0.021028 &  0.027981\\
  P  &  0.000090 &  0.000099\\
  S  &  0.008181 &  0.009055\\
  Cl &  0.000110 &  0.000122\\
  Ar &  0.000877 &  0.000970\\
  K  &  0.000037 &  0.000041\\
  Ca &  0.001238 &  0.001573\\
  Sc &  0.000000 &  0.000001\\
  Ti &  0.000039 &  0.000049\\
  V  &  0.000001 &  0.000002\\
  Cr &  0.000152 &  0.000169\\
  Mn &  0.000030 &  0.000037\\
  Fe &  0.011036 &  0.012214\\
  Co &  0.000032 &  0.000033\\
  Ni &  0.000504 &  0.000598\\
\hline
\end{tabular}
\end{table*}
                                          
\label{lastpage}                          

\end{document}